\documentclass[twocolumn,superscriptaddress,preprintnumbers,amsmath,10pt,aps,prl,nobibnotes,longbibliography]{revtex4-1} 

\usepackage{algorithm, algorithmic}

\usepackage{amsmath}
\usepackage{amsfonts}
\usepackage{bm}
\usepackage{color}
\usepackage{verbatim}
\usepackage{graphicx}
\usepackage{lipsum}
\usepackage[utf8]{inputenc}
\usepackage{times}
\usepackage{graphicx}
\usepackage{bm}
\usepackage{amssymb,multirow}
\usepackage[dvipsnames]{xcolor}
\usepackage{soul}
\usepackage[normalem]{ulem} 
\usepackage[mathlines]{lineno}
\usepackage{bbm}
\usepackage{siunitx}
\usepackage{array}
\usepackage{physics}

\def\Vecv{\mathbf{v}}

\def\VecI{\mathbf{I}}
\def\VecJ{\mathbf{J}}

\def\VecH{\mathbf{H}}
\def\VecM{\mathbf{M}}
\def\VecP{\mathbf{P}}

\def\VecS{\mathbf{S}}

\def\VecV{\mathbf{V}}

\def\VecX{\mathbf{X}}

\begin{document}

\title{High-dimensional spatial mode sorting and optical circuit design\\ using multi-plane light conversion}

\author{Hlib Kupianskyi}
\email{hk422@exeter.ac.uk}
\affiliation{Physics and Astronomy, University of Exeter, Exeter, EX4 4QL. UK.}
\author{Simon~A.~R.~Horsley}
\affiliation{Physics and Astronomy, University of Exeter, Exeter, EX4 4QL. UK.}
\author{David~B.~Phillips}
\email{d.phillips@exeter.ac.uk}
\affiliation{Physics and Astronomy, University of Exeter, Exeter, EX4 4QL. UK.}

\begin{abstract}
Multi-plane light converters (MPLCs) are an emerging class of optical device capable of converting a set of input spatial light modes to a new target set of output modes. This operation represents a linear optical transformation -- a much sought after capability in photonics. MPLCs have potential applications in both the classical and quantum optics domains, in fields ranging from optical communications, to optical computing and imaging.  They consist of a series of diffractive optical elements (the `planes'), typically separated by free-space. The phase delays imparted by each plane are determined by the process of inverse-design, most often using an adjoint algorithm known as the wavefront matching method (WMM), which optimises the correlation between the target and actual MPLC outputs. In this work we investigate high mode capacity MPLCs to create arbitrary spatial mode sorters and linear optical circuits. We focus on designs possessing low numbers of phase planes to render these MPLCs experimentally feasible. To best control light in this scenario, we develop a new inverse-design algorithm, based on gradient ascent with a specifically tailored objective function, and show how in the low-plane limit it converges to MPLC designs with substantially lower modal cross-talk and higher fidelity than achievable using the WMM. We experimentally demonstrate several prototype few-plane high-dimensional spatial mode sorters, operating on up to 55 modes, capable of sorting photons based on their Zernike mode, orbital angular momentum state, or an arbitrarily randomized spatial mode basis. We discuss the advantages and drawbacks of these proof-of-principle prototypes, and describe future improvements. Our work points to a bright future for high-dimensional MPLC-based technologies.
\end{abstract}

\maketitle

\noindent{\bf Introduction}\\
Light fields are represented as complex multi-dimensional vector functions that map the variation in intensity, phase and polarisation of each spectral component in space. These numerous degrees-of-freedom translate into a high information capacity, which is an attractive resource for a number of emerging applications in photonics. The technology to manipulate the spectral and polarisation dimensions of light is relatively mature (using e.g.\ prisms, diffraction gratings and polarising beam-splitters), however methods to address spatial degrees-of-freedom are still in their infancy. A key operation is the sorting of photons based on their transverse spatial mode~\cite{miller2019waves}. This is equivalent to performing a spatial change of basis operation, which can be physically accomplished by a {\it spatial mode sorter}: a device that redirects the energy carried by a set of orthogonal input spatial modes to separate locations across a transverse plane at its output. Spatial mode sorters are an example of more general optical transformations that can be understood as linear {\it optical circuits}, which map a group of input spatial modes to a new group of output modes: the action of an arbitrary optical circuit being described by a linear matrix operator.

Spatial mode sorters and optical circuits have a diverse range of applications, including, for example, employment as spatial multiplexing elements to expand the bandwidth of optical communication links~\cite{gibson2004free,berkhout2010efficient,bozinovic2013terabit,kahn2017communications,puttnam2021space}; unscrambling light to visualise objects hidden behind opaque media~\cite{popoff2010measuring,vcivzmar2012exploiting,butaite2022build}; far-field super-resolution imaging schemes~\cite{pushkina2021superresolution,bearne2021confocal}; quantum cryptography~\cite{bechmann2000quantum,mirhosseini2015high}, and use in emerging classical and quantum optical computation architectures~\cite{mair2001entanglement,carolan2015universal,leedumrongwatthanakun2020programmable,fickler2020full,lib2021reconfigurable,lib2022quantum}. The development of new hardware capable of operating on the spatial degrees-of-freedom of light is needed to unlock these new technologies.

\begin{figure*}[t]
   \includegraphics[width=1\textwidth]{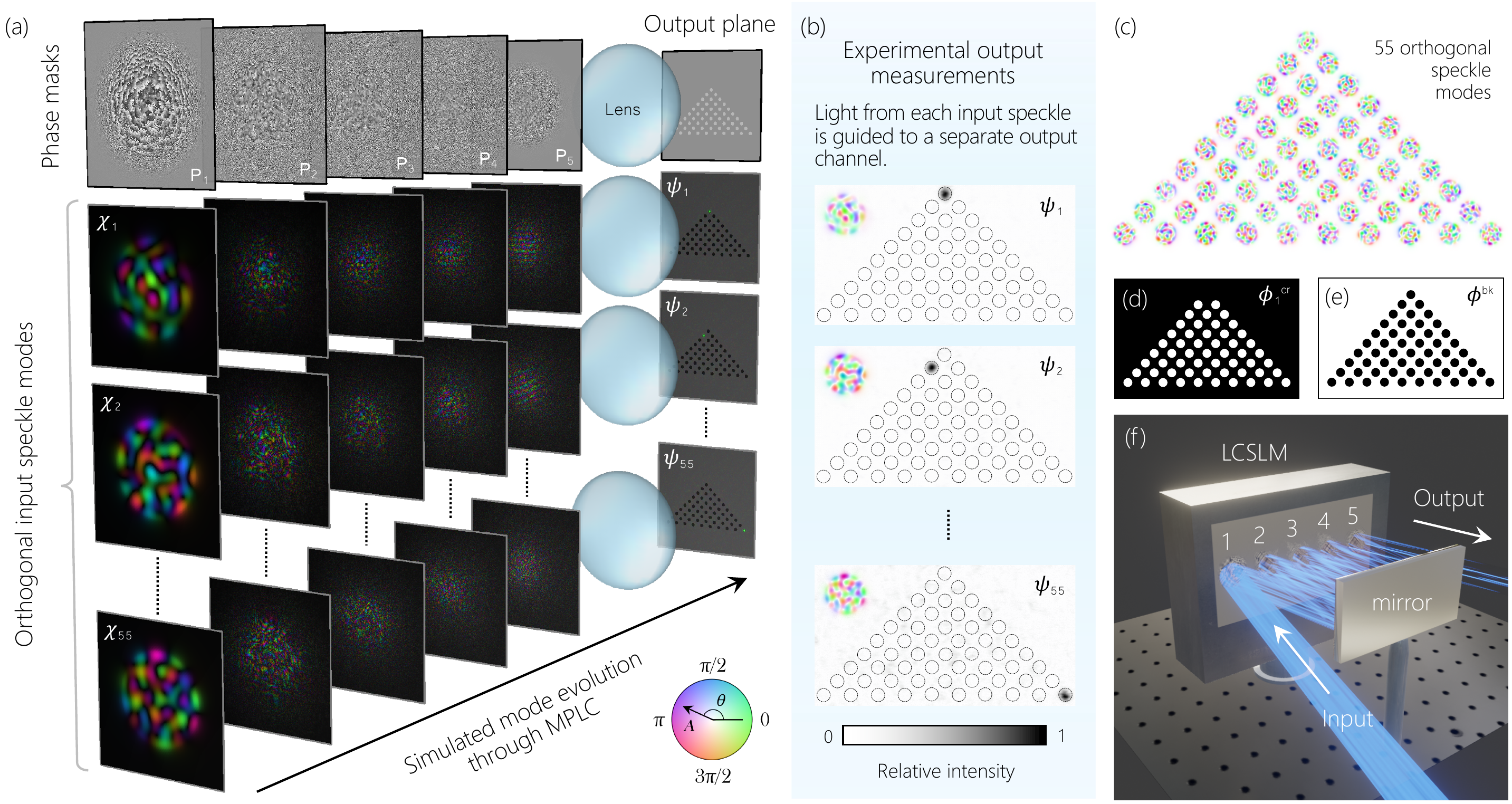}
   \caption{{\bf A high-dimensional MPLC-based spatial mode sorter}. (a) Simulated mode evolution through an optimised MPLC. The top row shows the intricate phase profiles of 5 phase masks ($\VecP_1$-$\VecP_5$) designed to spatially sort 55 orthogonal `speckle modes'. The phase profile of each mask is depicted on a greyscale colourmap to highlight the presence of many phase wraps between adjacent pixels. The triangular configuration of the 55 circular output channels is shown on the right-hand-side after a lens Fourier transforms the field to the output plane. The lower 3 rows in (a) show the evolution of 3 of the speckle modes through the MPLC ($\boldsymbol{\chi}_1$,$\boldsymbol{\chi}_2$ and $\boldsymbol{\chi}_{55}$). The fields arriving at each plane are displayed. The light from each speckle mode input is focussed into separate output channels -- we see the green foci appearing in different channels at the output plane. Here we display on a black background to ease visualisation the fields propagating through the MPLC. (b) Examples of experimentally measured output plane intensity measurements as the 3 speckle patterns depicted in (a) are injected into a prototype implementation of the mode sorter. (c) A rendering of all 55 orthogonal input speckle modes. (d) An example of the function $\boldsymbol{\phi}_1^{\text{cr}}$ which can be used to suppress cross-talk in Eqn.~\ref{Eqn:objective}. An example of $\boldsymbol{\phi}^{\text{bk}}$ which can be used tune output fidelity vs.\ efficiency in Eqn.~\ref{Eqn:objective}. In (d) and (e), white = 1 and black = 0. (f) A schematic of how our prototype MPLC is implemented using a liquid-crystal spatial light modulator (LCSLM) placed opposite a mirror, with the reflections numbered.}
   \label{Fig:concept}
\end{figure*}

Modal decomposition of monochromatic light fields can be performed, one mode at a time, by sequential projective measurements made using a dynamic spatial light modulator (SLM)~\cite{gibson2004free,flamm2012mode}. This technique has an overall efficiency that scales with $\sim1/N$ or lower, where $N$ is the size of the basis to be decomposed. However, spatial mode sorters are fundamentally different: they are capable of passively operating on all input modes simultaneously. This removes any fundamental limits on conversion efficiency or dependence on SLM modulation rate, and preserves the coherence of the input field at the output, which is crucial to many applications. Despite this promise, efficient optical transformation of large numbers of spatial modes (e.g.\ greater than $N\sim10$) is a challenging task, and a variety of different approaches are under development.

Over two decades ago, Reck et al.\ showed that any unitary transformation, which includes the operation of mode sorters and optical circuits, can be efficiently achieved using a network of beamsplitters and phase shifters~\cite{reck1994experimental}. However, the number of components required scales as ${N(N-1)/2}$, e.g.\ sorting $N=50$ modes would require over 1200 individual beamsplitters. Such networks are under development in the form of photonic integrated circuits -- meshes of waveguides on-chip~\cite{clements2016optimal}. These devices have exciting future potential, however stringent fabrication tolerances and inherent losses currently limit them to $N$\,$\sim$\,10 modes, and this bespoke technology is not yet widely available~\cite{bogaerts2020programmable}.

Multiple scattering materials, such as multi-mode fibres or thin layers of paint, have also been exploited as a mode transforming medium~\cite{huisman2015programmable,fickler2017custom,matthes2019optical,leedumrongwatthanakun2020programmable,defienne2020arbitrary}. By considering such media as a pre-fabricated high-dimensional random network, it is possible to identify and guide light through subsets of the medium that approximate a target optical transformation -- although the probability of finding suitable elements falls as $N$ increases, and so levels of crosstalk tend to grow rapidly with $N$. Furthermore, taking this approach usually results in a significant proportion of the input light being lost due to uncontrolled scattering.

A third category of optical circuit relies on diffractive elements which impart a spatially varying phase delay to incident light fields. A single diffractive element can be designed to sort arbitrary orthogonal spatial modes by multiplexing several diffraction gratings~\cite{gibson2004free,vcivzmar2012exploiting,fickler2020full}. However, each input mode diffracts from all of the multiplexed gratings. Therefore, as the number of modes $N$ grows, the efficiency of this approach falls steeply~\cite{mazilu2017modal}.

Allowing light to interact with several diffractive elements consecutively expands the sub-space of possible transformations that can be performed, and improves the efficiency~\cite{wang2017programmable}. Multi-plane light converters (MPLCs) are an emerging class of optical device that achieve this using a cascade of diffractive elements (the `planes') separated by free-space~\cite{morizur2010programmable,labroille2014efficient} -- see Fig.~\ref{Fig:concept}. They can be understood as an {\it artificially engineered scattering medium}~\cite{boucher2021full}, designed to achieve a target optical transformation, and coarse-grained into discrete planes to ease both design and implementation. Due to their versatility, MPLCs are attracting increasing interest, and are finding a growing number of applications across classical and quantum optics~\cite{wang2018dynamic,hiekkamaki2019near,brandt2020high,fickler2020full,mounaix2020time,lib2021reconfigurable}. For a few specific modal bases, elegant MPLC designs have been found that are capable of efficiently sorting many spatial modes using a small number of planes, see for example refs~\cite{berkhout2010efficient,fontaine2019laguerre,butaite2022build}. However, in the more general case of arbitrary basis transformations, all the studies to date have demonstrated relatively low-dimensional MPLCs, typically operating on less than $N\sim10$ spatial modes, with approximately a 1:1 ratio of modes:planes.

In this work we explore the high-dimensional spatial mode transforming capabilities of MPLCs. We design and experimentally demonstrate MPLCs capable of sorting up to 55 spatial modes of arbitrary basis using $N=5$ planes. Design of the phase delay profiles imparted to light interacting with each plane of an MPLC is an inverse problem: there is no known method to directly calculate the {\it optimal} phase profiles that achieve a given target optical transform in the simplest way (which is often understood to be the smallest number of planes possible). Therefore, iterative optimisation methods are typically employed.

We devote the first part of our paper to this inverse design problem. We develop a new design algorithm, based on gradient ascent with a carefully chosen objective function, and show that in the low-plane limit it converges to MPLC designs which outperform those generated using the state-of-the-art methods (e.g.\ the wavefront matching method~\cite{hashimoto2005optical}). In particular, we show that the modal cross-talk of a mode sorter -- the fraction of input light sent to the wrong output channels -- can be suppressed by up to an order of magnitude, and the fidelity of the output modes generated by an optical circuit can be substantially boosted. Furthermore, our approach enables the trade-offs between modal cross-talk, transform efficiency and output mode fidelity to be freely adjusted.

In the second part of our paper we focus on experimentally realising a range of prototype MPLCs. We demonstrate spatial mode sorters capable of separating randomly generated orthogonal speckle patterns, as well as new MPLC designs for sorting modes by their orbital angular momentum (OAM) and Zernike mode. We discuss the advantages and drawbacks of these proof-of-principle prototypes, and describe future improvements. Our work sheds new light on MPLC design schemes, and expands the toolbox of high-dimensional optical mode sorters, mode combiners, and photonic circuits for future imaging, communications and optical computing applications.\\

\noindent{\bf Multi-plane light conversion}\\
We begin by sketching out the MPLC design problem from a mode sorting perspective. We aim to sort $N$ orthogonal spatial light modes incident on the input of the MPLC, into separate output channels each defined by a Gaussian spot focused to a different lateral position. Figure~\ref{Fig:concept}(a) shows a schematic, where in this case we define the output to be in the Fourier plane of the last MPLC plane. We assume the polarisation of the input modes is uniform, and the optical systems are of low numerical aperture (NA), and so use the scalar diffraction theory in what follows.

The basis of modes to be sorted can be captured by matrix $\VecX \in \mathbb{C}^{P \times N} $, which transforms from the $N$-dimensional input mode basis to the $P$-dimensional pixelated real-space basis, where $P>>N$ thus the field is over-sampled in the pixel basis. Column $n$ of $\VecX$ is formed from the scalar real-space representation of the $n^{\text{th}}$ input spatial mode which has been sampled at the pixel resolution and reshaped into a column vector $\boldsymbol{\chi}_n$. The target (ideal) output basis of focused spots can be similarly defined by matrix $\boldsymbol{\Phi} \in \mathbb{C}^{P \times N}$.

The forward model capturing the operation of an MPLC consisting of $M$ planes with an arbitrary set of phase profiles is given by transfer matrix $\VecS \in \mathbb{C}^{P \times P}$, represented in the real-space input and output bases:
\begin{equation}\label{Eqn:MPLCreal}
    \begin{split}
    \VecS& = \VecP_M\cdot\VecH\cdot\VecP_{M-1}\dotsb\VecH\cdot\VecP_{2}\cdot\VecH\cdot\VecP_1\\
    & = \VecP_{M}\cdot\overset{\curvearrowleft}{\prod_{m=1}^{M-1}}\left[\VecH\cdot\VecP_m\right].
    \end{split}
\end{equation}
$\VecS$ describes how any input field will be transformed after propagation through the MPLC. Here $\VecP_m$ is a unitary diagonal matrix whose diagonal elements hold the phase change imparted by each pixel of the $m^{\text{th}}$ plane. $\VecH$ is a free-space propagation matrix transforming the field from the output facet immediately after plane $m$ to the input facet immediately before plane $m+1$. We note that equivalently $\VecH$ can also be modified to represent any other linear media placed between one phase plane and the next~\cite{goel2022inverse}.

In order to achieve the target mode sorting operation, $\VecS$ should satisfy
\begin{equation}\label{Eqn:dimRed}
    \boldsymbol{\Phi}^{\dagger}\cdot\VecS\cdot\VecX = \VecI,
\end{equation}
where $\VecI \in \mathbb{R}^{N \times N}$ is the identity matrix, and $(\cdot)^{\dagger}$ represents the conjugate transpose operator.
Equation~\ref{Eqn:dimRed} states that transforming the basis of $\VecS$ into the lower $N$-dimensional bases of the target input and output mode sets should yield the identity matrix: i.e.\ the input modes are perfectly transformed into the output modes with no cross-talk. Equation~\ref{Eqn:dimRed} also captures the more general process of converting any input basis to an arbitrary new output basis defined by choice of matrix $\boldsymbol{\Phi}$ -- the operation of an arbitrary optical circuit. As $\VecS$ is unitary, the inner product of any pair of input modes is preserved at the output.

Our task is to find the spatially varying phase delays that each plane should impart to a propagating field (i.e.\ the diagonal elements of $\VecP_i$), such that matrix $\VecS$ comes close to satisfying Eqn.~\ref{Eqn:dimRed} with a number of planes $M$ that is experimentally feasible -- and ideally as small as possible to reduce the complexity, bulk and loss of the final device. We may also wish to place addition constraints on $\VecS$, such as encouraging the phase profiles of the planes to vary smoothly, which reduces scattering losses and increases the spectral bandwidth over which the MPLC will operate~\cite{fontaine2019laguerre}. Note that Eqn.~\ref{Eqn:dimRed} constrains the relative phase of the output modes to zero. In some cases we may have no need to fix these relative output phases -- and this additional freedom may lead to a better mode sorting solution. In such cases the right-hand-side of Eqn.~\ref{Eqn:dimRed} is replaced with a unitary diagonal matrix with arguments that are allowed to take arbitrary values. We can qualitatively see from Eqn.~\ref{Eqn:MPLCreal} that as $M$ and $P$ grow, so does the sub-space of possible transforms $\VecS$ can take, thus improving the likelyhood of there existing, and of us finding, phase profiles that comes close to satisfying Eqn.~\ref{Eqn:dimRed}.

Determining the {\it minimum} number of planes required to sort or transform a given set of spatial modes is an open problem. A conceptually similar device to an MPLC has been theoretically proposed, that consists of a system of phase planes separated by segments of multimode waveguide. In this case it has been shown that ${M = 6N+1}$ planes are required to enact any unitary transformation on an arbitrary basis of $N$ input modes~\cite{pastor2021arbitrary}, a limit derived by mapping the operation onto that of a network of interferometers~\cite{clements2016optimal}. A similar limit is expected to apply to free-space MPLCs. Qualitatively, we might expect efficient arbitrary transformation of $N$ modes could be achieved in about $2N$ to $3N$ planes -- see Supplementary Information (SI) \S1 for discussion of this limit. 

The numbers of phase masks imposed by these limits are impractical for performing high-dimensional transforms on tens of spatial modes. However, depending on the application, some degree of loss, reduction in fidelity, and modal cross-talk may be tolerated. This situation corresponds to imperfectly satisfying Eqn.~\ref{Eqn:dimRed}, such that the change of basis operation on the optimised matrix $\VecS$ instead yields ${\boldsymbol{\Phi}^{\dagger}\cdot\VecS\cdot\VecX = \VecJ}$, where $||\VecJ-\VecI||$ is smaller than some agreed tolerance. It is this scenario we are interested in here, and we ask -- how best to design an MPLC in the low-plane limit?\\

\noindent{\bf MPLC design using gradient ascent}\\
In general, MPLCs can be engineered through the process of inverse design: an iterative procedure is used to find a set of phase profiles that approximate a target operation. Despite the large number of parameters to optimise (a total of $MP^2$ pixels), the {\it wavefront matching method} has proven to be an efficient way to solve this problem~\cite{hashimoto2005optical,wang2018dynamic,fontaine2019laguerre}. Wavefront matching is an adjoint algorithm that enables fast parallelised calculation of how the phase profile of each plane should be iteratively modified to improve the spatial overlap between the target outputs and actual outputs of the MPLC. See SI \S4 for a detailed description. However, while this has proven to be a powerful technique, it does not allow the target objective function to be freely chosen. We now explore if it is possible to do better with a more flexible design algorithm.

We cast the MPLC design problem in terms of gradient ascent -- a `hill climbing' optimisation algorithm that iteratively improves the design to maximise a target objective function~\cite{barre2022inverse}. We introduce a specifically tailored objective function to allow the trade-off between output mode fidelity, cross-talk and conversion efficiency to be adjusted -- important for the design of high-dimensional few plane MPLCs, where not all input light can be fully controlled. We aim to maximise the value of the real positive scalar $F_{\text{T}} = \sum_{n=1}^{N}F_n$, where $n$ indexes the $N$ input modes and
\begin{equation}\label{Eqn:objective}
    F_n = \underbrace{\alpha \left|\boldsymbol{\psi}^{\dagger}_n\cdot\boldsymbol{\phi}_n\right|^2}_\text{Fidelity - phase free} -\underbrace{\beta\;{\rm Re}\left[\boldsymbol{\psi}^{\dagger}_n\cdot\boldsymbol{\psi}^{\text{cr}}_n\right]}_\text{Cross-talk} +\underbrace{\gamma\;{\rm Re}\left[\boldsymbol{\psi}^{\dagger}_n\cdot\boldsymbol{\psi}^{\text{bk}}_n\right]}_\text{Efficiency}.
\end{equation}
The three contributions to our objective function can be understood as follows:

The first term on the right-hand-side (RHS) of Eqn.~\ref{Eqn:objective}, weighted by positive scalar $\alpha$, is designed to maximise the overlap between the 
$n^{\text{th}}$ target output mode represented by column vector $\boldsymbol{\phi}_n$ and the $n^{\text{th}}$ MPLC output, $\boldsymbol{\psi}_n=\VecS\cdot\boldsymbol{\chi}_n$.
Taking the absolute square of this overlap leaves the relative global phase of the light sorted into each output channel unconstrained. If necessary this term can be modified to enforce the relative phase between the outputs, as will be shown later.

The second term on the RHS of Eqn.~\ref{Eqn:objective} is weighted by the positive scalar $\beta$ and is designed to minimise cross-talk between the output channels.  For a mode sorter design, ${\boldsymbol{\psi}^{\text{cr}}_n = \boldsymbol{\psi}_n\odot\boldsymbol{\phi}_n^{\text{cr}}}$, where the operation $\odot$ signifies the element-wise Hadamard product, and $\boldsymbol{\phi}_n^{\text{cr}}$ defines the locations of the {\it wrong} output channels, i.e.\ its elements are set to 1 inside all output channels other than the target channel and to zero everywhere else -- Fig.~\ref{Fig:concept}(d) shows an example. This term tends to zero in the limit of no cross-talk, and here we subtract it so that non-zero levels of cross-talk act to decrease $F_n$.

Thirdly, we can also allow some of the light to be deliberately scattered into a designated background region outside all of the output channels. This lowers the efficiency of the device, but can enable solutions with higher fidelity and/or lower cross-talk to potentially be accessed. The positive scalar $\gamma$ weights the importance of this third term on the RHS of Eqn.~\ref{Eqn:objective}, and ${\boldsymbol{\psi}^{\text{bk}}_n = \boldsymbol{\psi}_n\odot\boldsymbol{\phi}^{\text{bk}}}$, where $\boldsymbol{\phi}^{\text{bk}}$ defines the background area surrounding all channels, i.e.\ its elements are set to zero inside any output channels, and to 1 everywhere else -- Fig.~\ref{Fig:concept}(e) shows an example. ${\boldsymbol{\psi}^{\dagger}_n\cdot\boldsymbol{\psi}^{\text{bk}}_n}$ is always positive, tending to zero in the limit when all transmitted light is directed within the boundaries of the output channels. Therefore adding this term acts to increase $F_n$, encouraging more light to be scattered around the output channels.

\begin{figure*}[t]
   \includegraphics[width=0.85\textwidth]{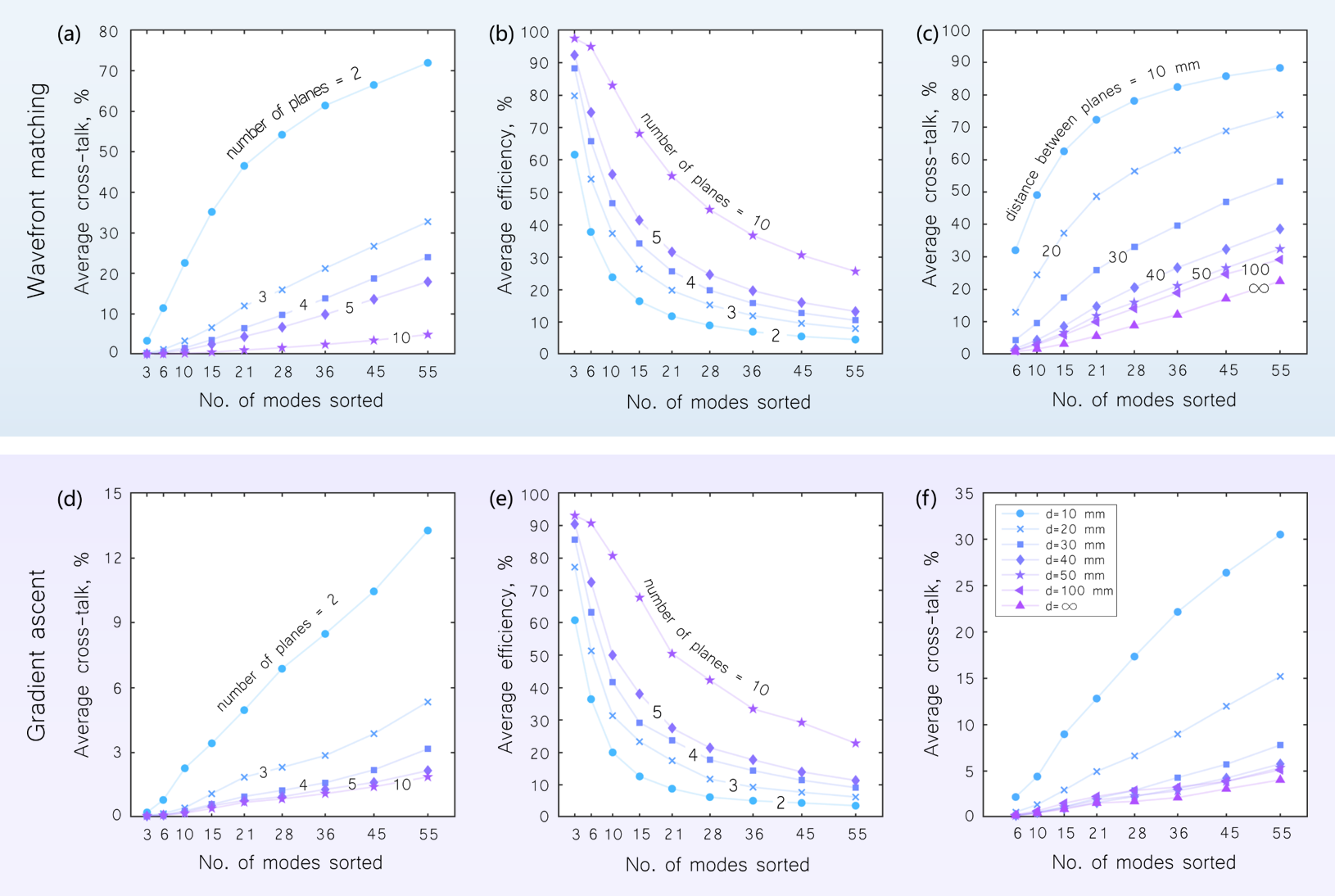}
   \caption{{\bf Simulated performance of mode sorter designs using wavefront matching vs.\ gradient ascent}. Top row, (a-c), shows the performance of designs optimised using the wavefront matching method. (a) Average cross-talk and (b) average mode sorting efficiency, as a function of modal capacity $N$ and number of planes $M$. (c) Average cross-talk as a function of modal capacity $N$ and plane separation, for a 5-plane MPLC. Here $\infty$ indicates a Fourier Transform between each plane, achievable with a lens. Bottom row (d-f), shows the equivalent graphs for designs optimised using our gradient ascent algorithm. Here $\alpha=1$, $\beta=2$, and $\gamma=0$ (i.e.\ chosen to maximise efficiency), and we do not constrain the relative phase of the output modes. Average cross-talk is significantly lower than (a) in all cases, while average sorting efficiency is roughly equivalent to (b). In these simulations, the wavelength is $633$\,nm, and each plane has a resolution of $700\times 700$ pixels, with a pixel pitch of 12.5\,$\mu$m -- chosen to match the typical pitch of pixels on liquid-crystal spatial light modulators. The distance between the phase planes is $\delta z=50$\,mm unless otherwise specified. }
   \label{Fig:comparison}
\end{figure*}

Our aim now is to derive an equation expressing how the phase profile on plane $n$ should be adjusted to improve our objective function. SI \S2 gives a detailed derivation. In summary, we examine the effect on the objective function $F$ due to a small variation of the $m^{\rm th}$ phase plane, $\VecP_{m}\to\VecP_{m}+\delta\VecP_{m}$. It is helpful to rewrite Eqn.~\ref{Eqn:MPLCreal}, the definition of the transfer matrix of the MPLC, as $\VecS=\VecS_{>}\cdot\VecP_{m}\cdot\VecS_{<}$, where $\VecS_{<,>}$ are respectively the sub-transfer matrices for propagation up to, and on from the $m^{\rm th}$ phase plane. Therefore, a change to the $m^{\rm th}$ phase plane leads to a modification in the transfer matrix ${\VecS\to\VecS+\delta\VecS}$, with ${\delta\VecS=\VecS_{>}\cdot\delta\VecP_{m}\cdot\VecS_{<}}$. Denoting the $n^{\rm th}$ input field as $\boldsymbol{\chi}_{n}$, this change in the transfer matrix leads to a change in the output field ${\delta\boldsymbol{\psi}_{n}=\delta\VecS\cdot\boldsymbol{\chi}_{n}}$, and thus a change in the partial objective function of
\begin{equation}\label{eq:deltaF}
    \delta F_{n}=2\;{\rm Re}\left[\frac{\partial F_{n}}{\partial\boldsymbol{\psi}_{n}}\cdot\delta\boldsymbol{\psi}_{n}\right]>0,
\end{equation}
which we demand to be positive to increase the value of our objective function. The gradient of $F_{n}$, with respect to output field $\delta\boldsymbol{\psi}_{n}$, is given by (see SI \S3)
\begin{equation}
    \frac{\partial F_{n}}{\partial\boldsymbol{\psi}_{n}}=\alpha\;\left(\boldsymbol{\psi}^{\dagger}_{n}\cdot\boldsymbol{\phi}_{n}\right)\boldsymbol{\phi}_{n}^{\dagger}-\tfrac{1}{2}\beta\;\boldsymbol{\psi}_{n}^{{\rm cr}\;\dagger}+\tfrac{1}{2}\gamma\;\boldsymbol{\psi}_{n}^{{\rm bk}\;\dagger}.\label{Eqn:objectiveGrad}
\end{equation}
Substituting for $\delta\boldsymbol{\psi}_{n}$ in Eqn.~\ref{eq:deltaF} yields,
\begin{equation}\label{eq:opt_eqn}
    \delta F_{n}=2{\rm Re}\left[\left(\frac{\partial F_{n}}{\partial\boldsymbol{\psi}_{n}}\cdot\VecS_{>}\right)\cdot\delta\boldsymbol{P}_{m}\cdot\bigg(\VecS_{<}\cdot\boldsymbol{\chi}_{n}\bigg)\right]>0.
\end{equation}
Equation (\ref{eq:opt_eqn}) has a straight-forward interpretation. It tells us that in order to improve the design of our MPLC for input mode $n$, we should propagate input field $n$ as far as the first boundary of the $m^{\rm th}$ phase plane and similarly back--propagate the corresponding partial derivative of the objective function (Eqn.~\ref{Eqn:objectiveGrad}) to the opposite boundary of the $m^{\rm th}$ phase plane. An incremental modification to this phase plane $\delta\VecP_{m}$ is then found by making a small phase change of fixed size $\delta\theta$ to each pixel, the sign of which is determined on a pixel-by-pixel basis to ensure that every contribution to Eqn.~\ref{eq:opt_eqn} is positive and thus increases the objective function.

To optimise the performance of the device for multiple modes simultaneously, changes required to improve the transformation of each individual mode can be averaged. We iteratively loop through all planes until the design converges. To ensure the sorting efficiency of each mode is equalised, if necessary at each iteration we adjust the relative importance of the gradient term associated with each respective mode to boost the importance of those modes that are currently sorted less efficiently.

When $\alpha = 1$ and $\beta = \gamma = 0$, our algorithm optimises the same objective as the wavefront matching method (when fixing the phase between outputs) -- in this special case the exact change to the phase profile of a single plane to optimally improve the objective function can be found (see SI \S4 and ref.~\cite{barre2022inverse}). We now show how mode sorting performance can be enhanced beyond solutions found using wavefront matching, when non-zero values of $\beta$ and/or $\gamma$ are introduced.\\

\begin{figure*}[t]
   \includegraphics[width=2\columnwidth]{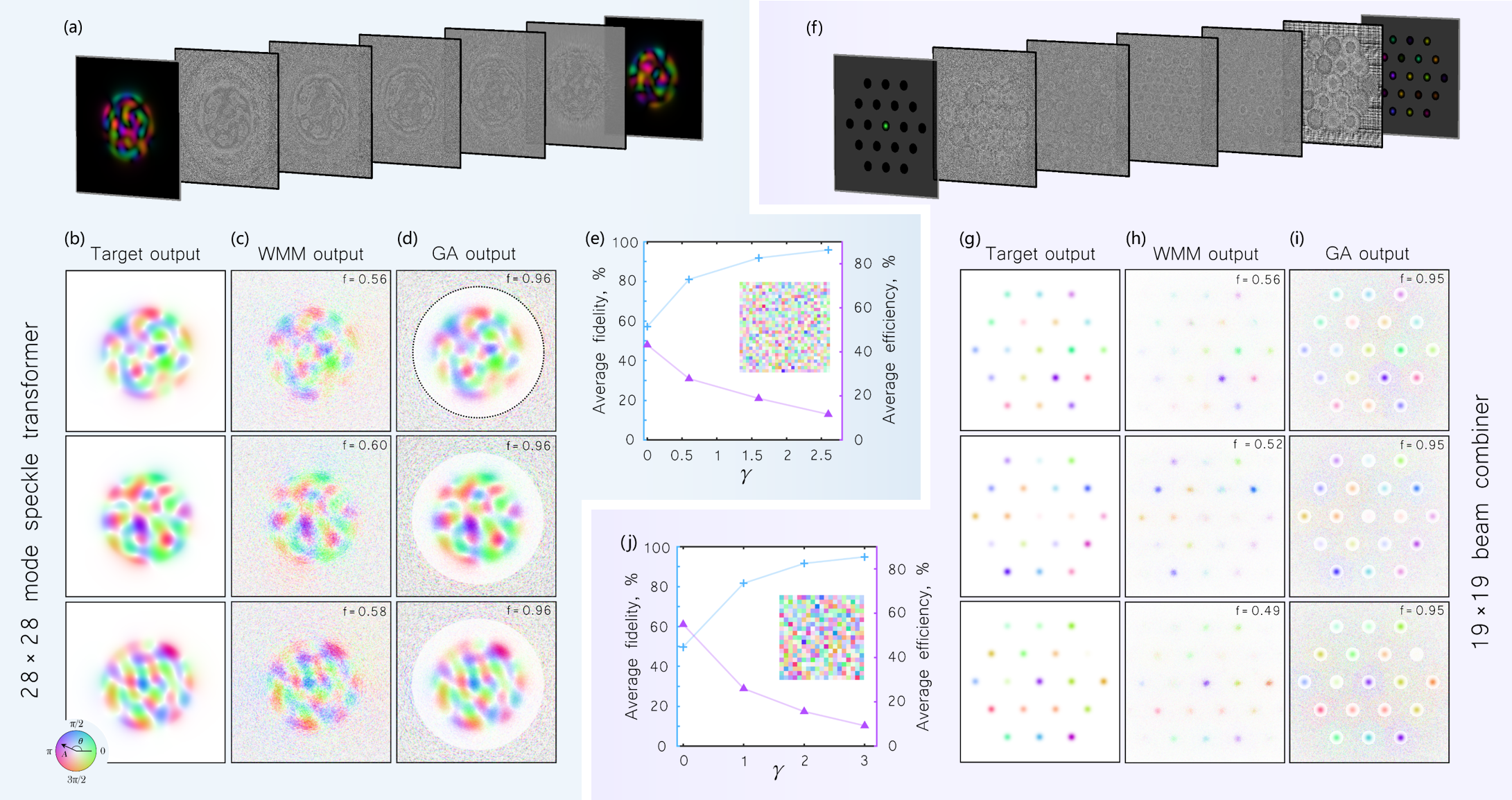}
   \caption{{\bf Arbitrary optical circuit design}. (a-e) Simulated performance of a 5-plane MPLC-based optical circuit designed to implement a randomly generated $28\times 28$ complex matrix transformation, acting on an input speckle mode basis. (a) Schematic of the {\it speckle mode transformer}, showing a single mode entering from the left and exiting on the right. (b) Three examples of the target output modes. (c) the corresponding modes generated by an MPLC designed using wavefront matching, and (d) gradient ascent ($\alpha = 1, \gamma = 2.6$). The black circle in the top row of (d) demarcates the boundary between the central single output channel and surrounding background region. Note that for display the output plane has been cropped close to this boundary. (e) Average output fidelity (blue crosses) and transform efficiency (purple triangles) as a function of $\gamma$. Inset shows the implemented $28\times 28$ random matrix transform. (f-j) Simulated performance of a 5-plane MPLC-based optical circuit designed to implement a randomly generated $19\times 19$ mode combining operation, acting on an input basis occupying a hexagonal array of Gaussian spots. (f) Schematic of the {\it beam combiner}, showing a single mode entering from the left and exiting on the right. The input and output channels have been highlighted. (g-i) Three examples of target output modes generated for this second optical circuit. In this case the background region is demarcated as anywhere outside the exclusion zones around the Gaussian spots. (j) Average output fidelity $f$ (blue crosses) and average transform efficiency $e$ (purple triangles) as a function of $\gamma$. Inset shows the implemented $19\times 19$ random matrix transform.}
   \label{Fig:OptCircuit}
\end{figure*}

\noindent{\bf Spatial mode sorting in an arbitrary basis}\\
We start by numerically bench-marking the capabilities of multi-plane light conversion to sort randomly generated orthogonal speckle patterns -- representing the most general sorting problem. Applications of such a sorter include, for example, unscrambling light that has propagated through opaque scattering media~\cite{vcivzmar2012exploiting,butaite2022build}. We investigate the sorting of between ${N=3-55}$ modes using ${M=2-10}$ planes (chosen to cover an experimentally realisable range). Examples of these input `speckle modes' are shown in Figure~\ref{Fig:concept}(c). Each speckle mode is formed from the complex weighted sum of a set of orthogonal step-index multimode fibre eigenmodes~\cite{li2021memory}, which ensures that the speckle modes are spatially localised. The speckle mode sets are constructed by randomly selecting the fibre-mode amplitudes (i.e.\ the complex weights) of the first speckle mode, after which the weights of subsequent modes are found using the Gram-Schmidt procedure.

Figure~\ref{Fig:comparison} shows a comparison of the numerically simulated cross-talk and efficiency of MPLC mode sorters designed using wavefront matching and our gradient ascent procedure (here with $\alpha = 1$, $\beta = 2$, $\gamma = 0$). See SI \S5 for the definitions of cross-talk and efficiency used here. For MPLCs optimised using wavefront matching, we find the average cross-talk varies approximately linearly with the size of the mode basis $N$ for $M>2$ planes, with performance improving as the number of planes increases as we would expect, as shown in Fig.~\ref{Fig:comparison}(a). In comparison, our gradient ascent algorithm generates MPLC designs with substantially lower levels of cross-talk, as shown in Fig.~\ref{Fig:comparison}(d). The enhancement is most striking for designs tasked with sorting high numbers of modes with low numbers of planes (i.e.\ lower ratios of $M/N$). For example, our simulations indicate that when sorting $N=55$ modes using $M=5$ planes, wavefront matching designs yield a cross-talk of $C_{\text{r}}\sim18\%$, which reduces by almost an order of magnitude to $C_{\text{r}}\sim2\%$ for designs optimised using gradient ascent.

Both design algorithms share similar efficiencies, which improve as $M$ increases and $N$ decreases, as shown in Figs.~\ref{Fig:comparison}(b,e). The separation between the planes also plays a role in MPLC performance. As the separation between planes is increased, light emanating from a single pixel on plane $m$ diffracts to illuminate a larger region of plane $m+1$. Increasing the level of inter-plane pixel-coupling in this way gives access to a wider range of possible solutions. Figure~\ref{Fig:comparison}(c,f) shows how the modal cross-talk of the optimised MPLC designs vary as a function of plane separation and mode capacity for an $M=3$ plane MPLC. We see that modal cross-talk decreases with increasing plane separation as we would expect. The performance of both algorithms can also be further improved by increasing the resolution of the masks. Here we used $700\times700$ pixels, representing a typical value that can be achieved using a liquid crystal spatial light modulator (LCSLM) to create each diffractive plane. For all plane separations we find our gradient ascent algorithm significantly outperforms the wavefront matching method in terms of modal cross-talk.

Over the range of modes and planes tested here, introducing a non-zero value of $\gamma$ only gave a small improvement in cross-talk at the expense of reduced efficiency. Control over the efficiency becomes more important when designing optical circuits, as we now show below.\\

\noindent{\bf High fidelity arbitrary optical circuit design}\\
Next we task our gradient ascent algorithm with the more general problem of designing arbitrary linear optical circuits -- see Fig.~\ref{Fig:OptCircuit}. In this case, the objective function associated with the $n^{\text{th}}$ mode becomes:
\begin{equation}\label{Eqn:objective2}
    F_n = \underbrace{\alpha\;{\rm Re}\left[\boldsymbol{\psi}^{\dagger}_n\cdot\boldsymbol{\phi}_n\right]}_\text{Fidelity - phase fixed}+\underbrace{\gamma\;{\rm Re}\left[\boldsymbol{\psi}^{\dagger}_n\cdot\boldsymbol{\psi}^{\text{bk}}_n\right]}_\text{Efficiency}.
\end{equation}
In Eqn.~\ref{Eqn:objective2}, the fidelity term has been modified to enforce the relative phase of the output modes. Secondly, as the output modes spatially overlap, the cross-talk term used in Eqn.~\ref{Eqn:objective} is no longer meaningfully defined and is therefore dropped. To tune the trade-off between fidelity and efficiency, at the output plane we demarcate the boundary between the output channel(s) and the background region where unwanted stray light can be shepherded (e.g.\ see black circle in Fig.~\ref{Fig:OptCircuit}(d), top row). The elements of $\boldsymbol{\phi}^{\text{bk}}$ are set to zero inside the region encompassing the output channel, and to 1 everywhere else.

We first design a 5-plane MPLC to accomplish a randomly generated complex matrix transformation, operating on an input basis formed from a set of $N = 28$ orthogonal speckle modes. The matrix transform is designed to generate a new set of speckle modes -- this {\it speckle transforming} operation is chosen to represent the most general form of optical circuit capable of acting on arbitrary input and output bases. Figure~\ref{Fig:OptCircuit}(a) shows a schematic of this design, with one example input and transformed output speckle mode. SI \S6 shows all of the 28 input and corresponding output modes this MPLC was designed for. Figure~\ref{Fig:OptCircuit}(b-d) shows a comparison of three of the simulated outputs generated by designs created using wavefront matching and gradient ascent. We design a range of MPLCs with $\alpha = 1$, and adjust the trade-off between average efficiency and output fidelity (defined in \S5) by tuning the value of $\gamma$ over the range $0<\gamma<2.6$, as shown in Fig.~\ref{Fig:OptCircuit}(e). When $\gamma = 0$ (i.e.\ equivalent to using the wavefront matching method), we find an average output fidelity of $f\sim0.58$. As $\gamma$ is increased, we see the output fidelity is improved at the expense of efficiency, in this case rising to a plateau of $f\sim0.96$. By redirecting the majority of unwanted stray speckle into the designated background region of the output, it can be blocked and prevented from polluting the optical system down-stream of the optical circuit -- a feature not possible with designs created using the wavefront matching method.

Another useful measure of optical circuit performance is the fidelity of the accomplished matrix transform itself, $f_{\text{M}}$ (see SI \S5 for definition). In this case $f_{\text{M}} = 0.95$ for the wavefront matching MPLC design, which rises to $f_{\text{M}} = 0.99$ for the highest fidelity gradient ascent MPLC design.

Figure~\ref{Fig:OptCircuit}(f) shows a second example of an optical circuit. Here we target the reverse operation to a mode sorter -- i.e.\ a {\it mode combiner}~\cite{butaite2022build}, which in this case uses 5 phase planes to transform 19 initially spatially separated Gaussian modes into a set of spatially overlapping modes -- each spread over a hexagonal array of Gaussian modes. SI \S6 shows all of the input and corresponding output modes. Confining both inputs and outputs to arrays of Gaussian spots would allow integration of this design with single mode waveguide arrays.

Due to the reciprocity of Maxwell's equations, simply transmitting light through a pre-designed low-plane mode sorter in the reverse direction does not lead to optimal performance of the mode combiner. This is because, when operating the device in the forward direction as a mode sorter, some light is scattered around the output channels. However, when the {\it ideal} sorted modes are transmitted through the device in the reverse direction to create a beam combiner, this scattered light is absent, thus reducing the fidelity of the beam combining operation. This problem can be overcome by specifically designing a beam combiner and tuning the objective function to shepherd unwanted light into the region around the combined beams.

We demonstrate this principle by designing a series of beam combiners using the objective function Eqn.~\ref{Eqn:objective2}, and as before, observe a similar trade-off between average fidelity and efficiency as $\gamma$ is increased (see Fig.~\ref{Fig:OptCircuit}(j)). Figure~\ref{Fig:OptCircuit}(g-i) shows a comparison between three of the target output modes generated using wavefront matching versus gradient ascent. The corresponding fidelity of the matrix transforms are $f_{\text{M}} = 0.77$ for the wavefront matching MPLC design, rising to $f_{\text{M}} = 0.94$ for the highest fidelity gradient ascent MPLC design.

In summary, we have shown that using gradient ascent with a careful choice of objective function may generate optical circuit designs that significantly outperform those created using wavefront matching in the low-plane limit. These simulations indicate that MPLC technology is a versatile and tunable platform to create high-dimensional and high-fidelity linear optical circuits using a feasible number of phase planes.\\

\begin{figure*}[t]
   \includegraphics[width=1.9\columnwidth]{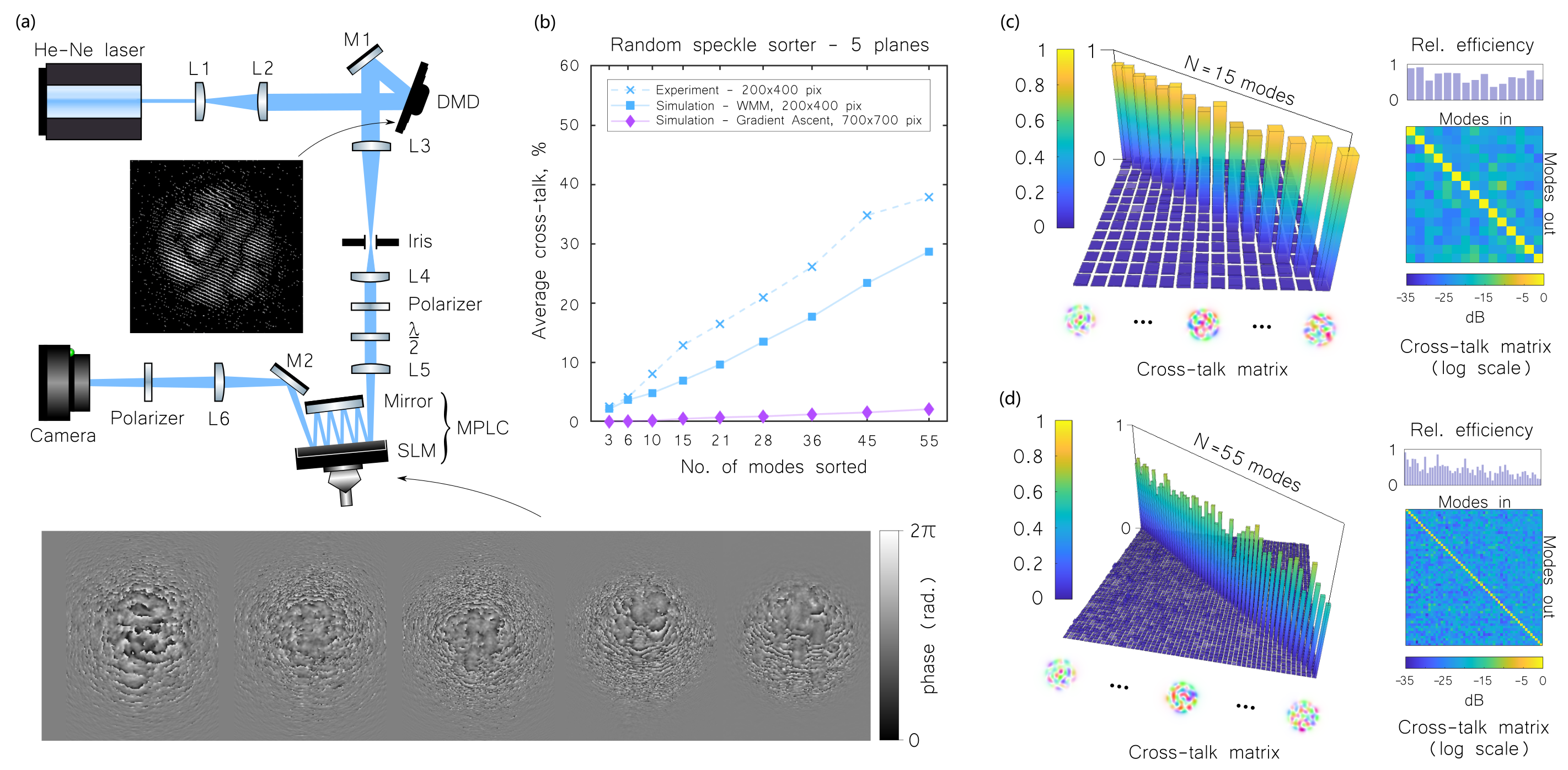}
   \caption{{\bf Prototype experimental implementation}. (a) A schematic showing the optical setup for a 5-plane MPLC system used to sort 55 orthogonal speckle mode in our prototype experiment. Insets show examples of a binary amplitude pattern displayed on the DMD to encode an input speckle mode, and a phase pattern displayed on the LCSLM operating as a 5-plane MPLC. (b) The experimentally measured levels of cross-talk vs.\ simulated levels of cross-talk for designs using wavefront matching (WMM) and our gradient ascent algorithm at higher resolution. (c,d) Experimental cross-talk matrices for (c) the 5-plane 15 speckle mode sorter and (d) the 5-plane 55 speckle modes sorter. Left-hand 3D plots show a linear scale; right-hand heatmaps shows the same data on a logarithmic scale. The relative efficiency with which each mode is sorted is also shown.}
   \label{Fig:setup_speckle}
\end{figure*}

\noindent{\bf Realisation of prototype high-dimensional mode sorters}\\
We now investigate the experimental implementation of high-dimensional spatial mode sorters using multiple reflections from a single liquid-crystal spatial light modulator (LCSLM). Figure~\ref{Fig:setup_speckle}(a) shows a schematic of our experimental setup. We use a digital micro-mirror device (DMD) to shape the input beams sent into the prototype mode sorter~\cite{mitchell2016high}. We test a series of 3-plane and 5-plane MPLCs by designing a range of devices enabling the sorting of between 3 to 55 modes in a variety of bases.

\vspace{1mm}

\noindent{\it Speckle mode sorting}\\
Movie 1 shows the recorded outputs of a 5-plane, 55 speckle-mode sorter as it is sequentially illuminated with each input speckle pattern in turn -- showing the successful sorting of 55 orthogonal speckle modes. Example outputs are shown in Fig.~\ref{Fig:concept}(b). Figure~\ref{Fig:setup_speckle}(b) shows how the level of experimentally measured cross-talk varies with number of sorted modes (crosses), in comparison to two simulated cases: (i) the cross-talk levels for a device with matching parameters to our experiment, designed using the objective parameters ${\alpha=1, \beta=\gamma=0}$ (squares); and (ii) cross-talk levels in a `best case' scenario, with higher resolution masks designed using parameters ${\alpha=1, \beta=2, \gamma=0}$ (diamonds), which we are unable to experimentally implement due to the limited resolution of our LCSLM. We include this simulation here to showcase the future capabilities of these devices. Figure~\ref{Fig:setup_speckle}(c,d) show examples of the experimentally measured cross-talk matrices and sorting efficiency when sorting ${N=15}$ (c) and ${N=55}$ (d) modes. The experimentally measured variation in sorting efficiency is due to differing levels of uncontrolled scattering experienced by different modes as they pass through the MPLC. Below we discuss future improvements to reduce such variation.

For the sorting of up to 6 modes, the experimental cross-talk levels match closely to those expected from our simulations (see Fig.~\ref{Fig:setup_speckle}(b)). However, beyond 6 modes, the experimental levels of cross-talk creep higher than those simulated. There are several possible reasons for this discrepancy. One challenge in the implementation of these high-dimensional MPLCs is the experimental alignment -- an issue exacerbated for larger numbers of sorted modes $N$ since the complexity of the phase masks increases with $N$. There are many alignment degrees-of-freedom that must be simultaneously optimised to obtain good performance: an MPLC consisting of $M$-planes requires the optimisation of $D\sim2M+4$ degrees-of-freedom, including pixel-perfect alignment of each plane -- see SI \S7 for more discussion. For example, $D\sim14$ for a 5-plane MPLC. To minimise the impact of alignment issues, we implemented a genetic algorithm that automatically optimised MPLC alignment (also carried out in ref.~\cite{brandt2020high}), once manual alignment had been conducted. \S7 gives more detail of this procedure.

The measured cross-talk will also incorporate the effects of imperfectly structured input beams, caused for example by the non-flat nature of the DMD screen on the scale of the wavelength. To mitigate this, we measured the aberrations in the optical system prior to the mode sorter, and apply a corrective phase function incorporated directly into the beam shaping patterns displayed on the DMD~\cite{vcivzmar2010situ,turtaev2017comparison}. 

Therefore, we believe the main reason for the higher levels of experimentally measured cross-talk, is the coupling between adjacent pixels in the displayed LCSLM patterns. The orientation of the liquid crystal layers are influenced by both the electric field addressing each pixel, and the orientation of the liquid crystal in neighbouring pixels, thus reducing the fidelity of the phase patterns that can be displayed~\cite{moser2019model,pushkina2020comprehensive}. Slowly spatially varying phase patterns -- such as those generally required for sorting low numbers of modes (e.g.\ $N\leq6$) -- are more faithfully reproduced than the more intricate phase patterns needed to sort higher numbers of modes. This problem is particularly severe at phase-wrapping boundaries which exhibit large jumps in liquid crystal orientation.

In our experimental study, we measured the lowest level of cross-talk for designs with $\alpha = 1$, and $\beta = \gamma = 0$ in the objective function given by Eqn.~\ref{Eqn:objective} (i.e.\ a design equivalent to the WMM). Best results were obtained by also including a smoothing constraint to the design algorithm to further suppress the effects of LCSLM pixel cross-talk, achieved following the strategy employed in ref.~\cite{fontaine2019laguerre} -- see SI \S4.
We also tested mode-sorters designed with parameters $\alpha = 1$, $\beta = 2$ and $\gamma = 0$, which in theory offer significantly lower levels of modal cross-talk (Fig.~\ref{Fig:comparison}). However, in general we found that these MPLC designs exhibited yet more complicated phase patterns, which were less faithfully represented when displayed on our LCSLM prototype, thus resulting in similar or higher levels of experimentally measured cross-talk.

Nonetheless, we consider the more intricate designs revealed by our simulations to be highly promising, and see several routes to successfully implementing them: using lithographically etched phase masks~\cite{fontaine2018packaged}, or metasurfaces~\cite{oh2022adjoint} in place of LCSLMs will eliminate pixel coupling, albeit at the expense of switchable operation. Alternatively, implementing each phase plane on a separate LCSLM would allow a larger number of pixels to be devoted to each plane than in our study, meaning the effect of pixel cross-talk could be minimised. Therefore, while here we prototyped these devices using a single LCSLM, we anticipate significantly higher levels of performance driven by our gradient ascent optimisation objective will be possible in the near future -- enhancing MPLC technology for a multitude of applications.

To illustrate the versatility of high-dimensional multi-plane light conversion, in the final part of our paper we investigate the sorting of some specific spatial mode sets that have a range of emerging applications: the separation of photons by their orbital angular momentum (OAM) state, and by Zernike spatial mode.\\

\begin{figure}[t]
   \includegraphics[width=1.0\columnwidth]{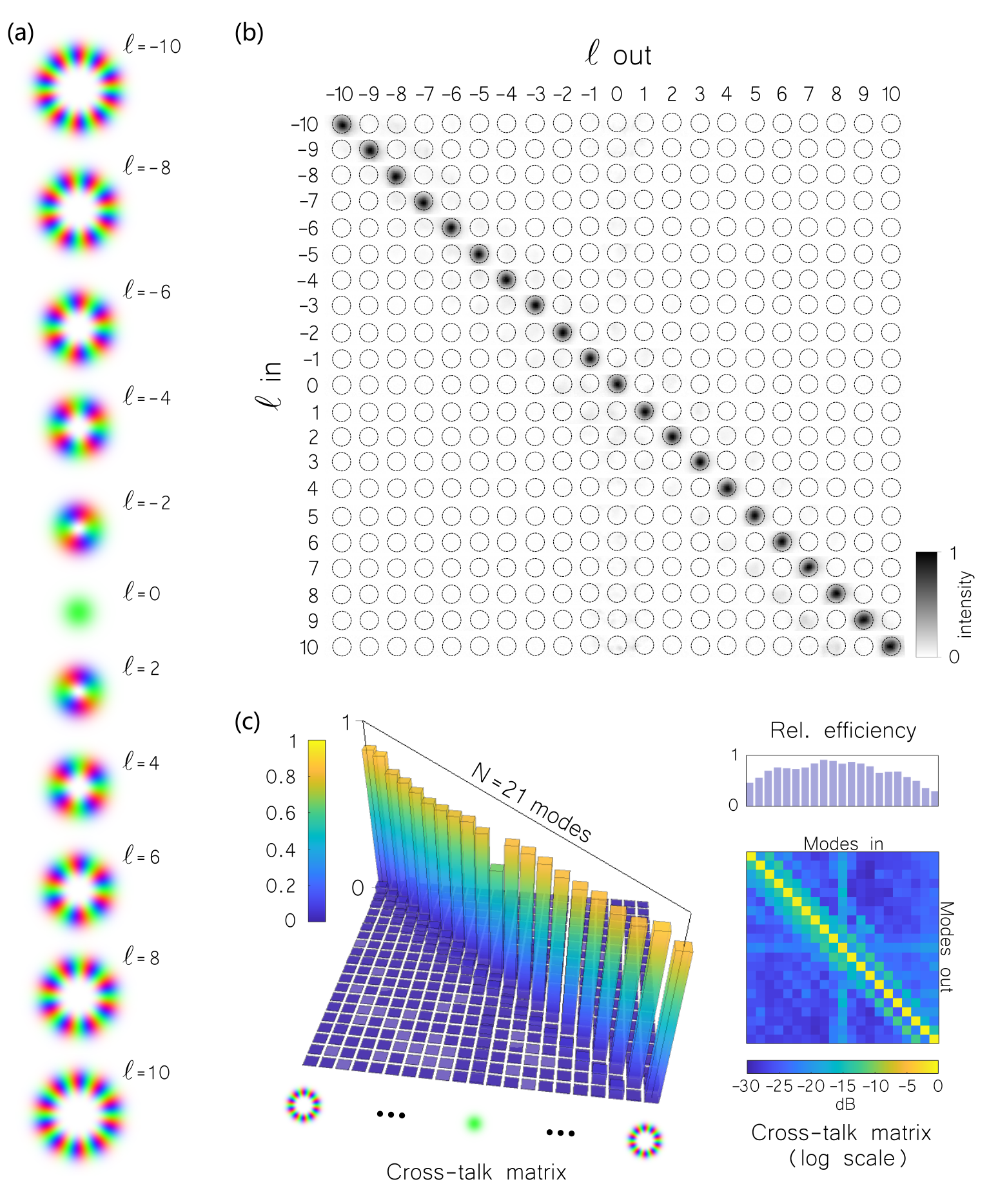}
   \caption{{\bf Prototype 3-plane 21 mode OAM sorter}. (a) Examples of LG modes of differing vortex charge. (b) Experimentally measured OAM sorter outputs (Gaussian spots). Each row shows the same rectangular letter box-shaped region-of-interest on the camera, as each of the 21 OAM states are transmitted through the OAM sorter, and the light is redirected into a laterally displaced output channels (demarcated by black circles). (c) Experimental cross-talk matrices, and relative efficiency bar graph for the prototype 3-plane 21 OAM sorter.}
   \label{Fig:oam}
\end{figure}

\noindent{\it OAM state sorting}\\
Beams carrying an integer value of orbital angular momentum (OAM), quantified by vortex charge $\ell$, are characterised by helical wavefronts. Examples are shown in Fig.~\ref{Fig:oam}(a). Spatially separating photons based on their OAM has attracted much interest in recent years~\cite{padgett2017orbital}. A 2-plane OAM mode sorter was first implemented using an optical Cartesian to cylindrical coordinate transform~\cite{berkhout2010efficient}. This device has an inherent average modal cross-talk of $\sim22\%$, which has been improved by incorporating beam copying elements~\cite{mirhosseini2013efficient} or by using a spiral transform~\cite{wen2018spiral}. These initial designs do not deliver control over the shape of the separated output beams -- limiting their utility for integrated applications. Recently, a 3-plane device solved this issue by reshaping outputs into Gaussian spots using a specially designed elliptical lens~\cite{wen2020compact}. Given the existence of these analytical, high-dimensional OAM mode sorting designs, it is interesting to benchmark how well our gradient ascent inverse design protocol performs this task. We ask: can we use inverse design to find other 3-plane solutions?

To test this, we design a 3-plane OAM sorter that separates input Laguerre-Gaussian (LG) modes (which each carry a well-defined vortex charge $\ell$) based on their OAM, and simultaneously focusses these outputs into Gaussian spots. We find the theoretical cross-talk of our designs increased approximately linearly with $N$, reaching $C_{\text{r}}\sim4$\,\% for $N=31$ LG modes (with vortex charges ranging between $\ell=\pm15$). In these designs we used our gradient ascent optimiser with $\alpha = 1, \beta = 2, \gamma = 0$. See SI \S8 for more detail. Unlike the analytical OAM sorter designs, these numerically optimised solutions fragment each input beam in an unrecognisable manner as it propagates through the MPLC. Nonetheless, we find that they have theoretical average levels of cross-talk on-par with state-of-the-art analytical methods ~\cite{mirhosseini2013efficient,wen2020compact}. Increasing the number of planes, spacing between planes and resolution would further improve performance.

Such numerically optimised OAM sorters may offer advantages in some scenarios. For example, the derivation of analytical OAM sorters are underpinned by the Paraxial approximation and the stationary phase approximation. Numerical optimisation is free from these approximations, which may benefit future ultra-compact sorter designs~\cite{oh2022adjoint}. Furthermore, using optimisation also enables designs to be specifically tuned for a target range of vortex charges and radial mode profiles.

\begin{figure*}[t]
   \includegraphics[width=1.8\columnwidth]{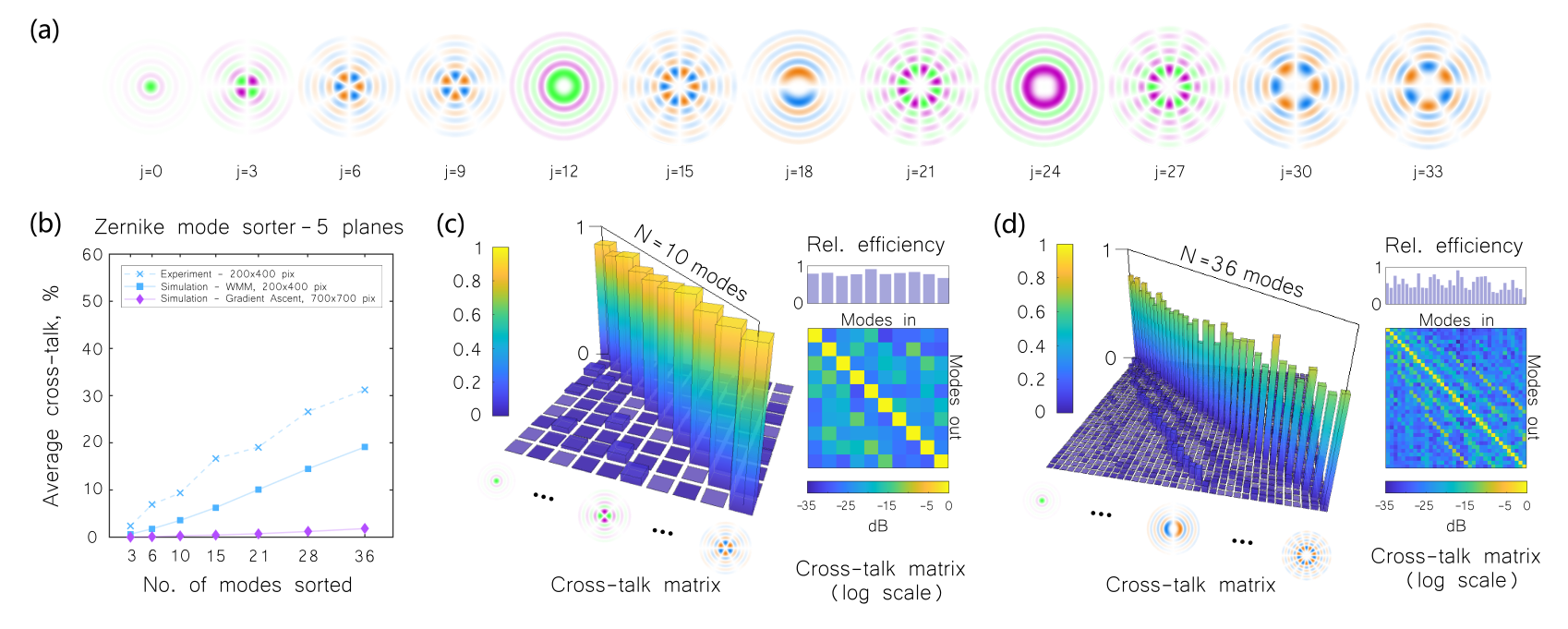}
   \caption{{\bf Zernike mode sorting}. (a) Examples of the Zernike modes (in this case their Fourier transforms) incident on the prototype MPLC, indexed using the OSA/ANSI notation: $j=\tfrac{1}{2}(n^2+2n+m)$. (b) Experimental average cross-talk levels when sorting between 3 and 36 Zernike modes using 5 planes, compared to the simulated performance of the wavefront matching method (WMM) and our gradient ascent algorithm at higher resolution. (c,d) Experimental cross-talk matrices and relative sorting efficiency for (c) the prototype 5-plane 10-mode Zernike sorter (average cross-talk $C_{\text{r}} \sim$ 9.4\%), and (d) the prototype 5-plane 36-mode Zernike sorter (average cross-talk $C_{\text{r}} \sim$ 31.2\%).}
   \label{Fig:Zern}
\end{figure*}

Movie 2 and Fig.~\ref{Fig:oam}(b-c) show the performance of a prototype experimental implementation of our 3-plane OAM mode sorter, in this case designed to separate 21 incident LG beams (vortex charges ranging through $\ell\pm10$). We use $\alpha=1$, $\beta=\gamma=0$ as before. The mean cross-talk is $C_{\text{r}}\sim19\%$, which can be compared to a simulated cross-talk of this low resolution design of $C_{\text{r}}\sim 9.5\%$. As all 3 planes of our prototype were encoded into a single LCSLM, we expect higher resolution devices will improve future performance.

As an extension, we also investigated the sorting of Bessel beams by both their vortex charge and radial index. Intriguingly, we found that our sorter works to some extent even for input modes outside the range it was designed for -- generating focussed spots at new output channel locations. Here we hypothesise that our optimised solution has settled close to a more general solution capable of sorting Bessel beams, inheriting an underlying symmetry of the basis in question. It may be interesting to explore this in future. SI \S8 gives more details.

We highlight that by increasing the number of planes to $\sim$7, Fontaine et al.\ found an elegant MPLC design capable of sorting hundreds of LG modes by both vortex charge and radial index~\cite{fontaine2019laguerre} -- thus sorting light fields of a complete basis~\cite{zhao2015capacity}. OAM and LG mode sorting are unusual examples whereby a large numbers of modes can be sorted with high efficiency using a relatively low number of planes -- something that is not possible for an arbitrary basis.\\

\noindent{\it Zernike mode sorting}\\
Finally, we explore the sorting of Zernike modes using multi-plane light conversion. The Zernike modes are a set of orthogonal functions which are typically associated with low order aberrations in optical systems -- see Fig.~\ref{Fig:Zern}(a), and SI \S8. Zernike mode sorting has been proposed as a route to achieve super-resolution confocal microscopy~\cite{bearne2021confocal}, or as an advanced spatial filter in exoplanet hunting telescopes, to separate the bright light of a parent star from the dim light emanating from an orbiting exoplanet~\cite{foo2005optical,carpenter2020laguerre}. 
Here we design and implement a range of 5-plane Zernike mode sorters, capable of sorting between 3 and 36 Zernike modes. Movie 3 shows the experimentally measured outputs of the device designed to sort 36 modes. Figure~\ref{Fig:Zern}(b) shows how the modal cross-talk scales with the number of modes, both experimentally and theoretically. We see that our gradient ascent algorithm yields low levels of modal cross-talk, and our experiments follow the same pattern as described above -- Fig.~\ref{Fig:Zern}(c,d) show more detail. To the best of our knowledge, these are the first theoretical or experimental demonstrations of Zernike mode sorting MPLCs.\\

\noindent{\bf Conclusions}\\
In summary, we have described a new way to design low-plane MPLCs, using gradient ascent with a carefully selected objective function. This enables the relative importance of output mode fidelity, mode sorting cross-talk, and transform efficiency to be tuned. We have also shown how the relative phase of output modes can be constrained or left as a free parameter in the design. Using this method we have discovered new MPLC designs for mode sorters that exhibit significantly lower modal cross-talk than designs found using the state-of-the-art wavefront matching method. We have also shown how our design algorithm can be tailored for high-fidelity mode combining and for the creation of general linear optical circuits. Our method allows for any differentiable objective function to be employed in MPLC design -- and SI \S2-4 provides extensive detail and examples of how to implement this strategy.

We note that gradient ascent/descent has been previously used to design the phase profiles of single diffractive phase masks, for spectral light shaping~\cite{xiao2016diffractive}, and 3D beam shaping~\cite{zhang20173d}. Our work highlights the potential of extending these devices to multiple planes. Very recently, gradient descent was also applied to the design of multi-mode graded-index devices~\cite{barre2022inverse} -- equivalent to the limit of very high numbers of closely-spaced smooth phase planes. In contrast, we have explored the opposite extreme, limiting the plane count in order to keep devices as simple to build as possible. In this regime we cannot fully control all input light, and so have developed new multi-parameter objective functions designed to suppress modal cross-talk and divert uncontrolled light into target areas where it can be blocked -- essential to achieve high fidelities in this scenario. These objectives functions may also prove useful when designing high mode capacity graded-index devices.

We have experimentally demonstrated MPLCs capable of sorting up to 55 randomly generated spatial modes, a new form of OAM sorter, and the first example (to our knowledge) of a Zernike mode sorter. SI \S7 provides a detailed explanation of how to align such devices. Our MPLC designs typically consist of rapidly spatially varying phase profiles which makes them difficult to accurately display on a single LCSLM. Nonetheless, these designs may be implemented in the future by explanding the resolution of each plane using several consecutive LCSLMs, or employing lithographically etched diffraction gratings or metasurfaces.

It is important to note one caveat of our approach -- the complexity of the phase patterns found in such low-plane high-dimensional MPLCs may limit their spectral bandwidth -- since smoothly varying phase profiles are required to ensure a spectral response which changes slowly with frequency~\cite{fontaine2019laguerre,butaite2022build}. However, we believe these methods are highly promising, and provide one of the only routes available to efficiently manipulate several tens to hundreds of spatial modes simultaneously. Through simulations and experiments we have shown that MPLCs have the capability to deliver optical functionality currently unmatched by any other platform.\\

\noindent{\bf Acknowledgements}\\
SARH acknowledges the Royal Society and TATA for financial support through grant URF\textbackslash R\textbackslash 211033. DBP thanks the Royal Academy of Engineering and the European Research Council (Grant no. 804626) for financial support.\\

\noindent{\bf Contributions}\\
DBP conceived the idea for the project. SARH derived the gradient ascent MPLC design method, and all the authors formulated the objective functions. HK performed all simulations and experiments, with supervision from DBP and SARH. All authors contributed to the analysis of the results and the writing of the paper.

\bibliography{fewModeLightConverterRefs}


\onecolumngrid
\newpage

\section*{Supplementary Information}

\noindent{\bf \S1: Qualitative analysis of the MPLC mode-to-plane scaling}\\\label{SI_qualitative_analysis}
\noindent Here we describe a qualitative argument to estimate the number of planes required to transform $N$ spatial modes with unit efficiency. 2 phase planes, with the second in the Fourier plane of the first, can approximately reshape a single input spatial mode in an arbitrary way using the well-known Gerchberg-Saxton (GS) algorithm~\cite{gerchberg1972practical,yang1994gerchberg}. In this case the phase profile of the first plane is first optimised using the GS algorithm to create the intensity profile of the target mode projected onto the second plane (with no constraint on the output phase profile). The phase profile of the second plane can then be chosen to correct the phase of the output to ensure it matches the target mode. Next consider sorting 2 input modes (labelled mode 1 and mode 2) using 4 planes: the first two planes can be assigned to arbitrarily transform mode 1, and the second two planes assigned to transform mode 2. In this case the presence of the phase planes assigned to the alternate mode modify the transformation that each set of phase planes must perform -- so they must all be optimised together using, for example, the gradient ascent methods present in the main paper. Since 2-planes can arbitrarily transform a single mode, it follows that 4 planes should be able to sort 2 input modes. This argument can be extended to approximate the target transformation of $N$ modes in roughly $M\sim2N$ planes. To ensure this transformation is high fidelity, we can assign 3 planes to transform each mode -- as it has recently been shown that 3 phase masks can achieve a near-perfect arbitrary transformation of a single input~\cite{hiekkamaki2019near}. Therefore using the above argument we would expect $N$ modes to be perfectly transformed using between $2N$ to $3N$ phase planes.\\

\noindent{\bf \S2: Derivation of gradient ascent-based MPLC design algorithm}\\\label{SI_GA}
As described in the main text, we aim to design the phase profiles of a series of $M$ phase masks to sort/transform $N$ spatial modes. To optimise the profiles of these phase planes, we define a real-valued objective function $F_{\text{T}} = \sum_{n=1}^{N}F_n$, where $n$ labels the set of spatial modes. $F_n$ is chosen to be some function of the complex output vector $\boldsymbol{\psi}_{n}$, and its conjugate transpose, i.e.\ $F_n(\boldsymbol{\psi}_{n},\boldsymbol{\psi}_{n}^\dagger)$ in the general case -- this functional form ensures $F_n$, and thus $F_{\text{T}}$, is a real-valued scalar as required (some specific examples are given later).

Changing the $m^{\rm th}$ phase mask by a small amount, $\VecP_{m}\to\VecP_{m}+\delta\VecP_{m}$ induces a change in the transfer matrix
\begin{align}
    \delta\VecS&=\left(\VecP_{M}\cdot\VecH\dots\VecP_{m+1}\cdot\VecH\right)\cdot\delta\VecP_{m}\cdot\left(\VecH\cdot\VecP_{m-1}\dots\VecH\cdot\VecP_{1}\right)\nonumber\\[5pt]
    &=\VecS_{>}\cdot\delta\VecP_{m}\cdot\VecS_{<}.
\end{align}
This change in the transfer matrix induces a corresponding change in the output fields of $\delta\boldsymbol{\psi}_{n}=\delta\VecS\cdot\boldsymbol{\chi}_{n}$, which changes the objective function as follows
\begin{align}
    \delta F_{\text{T}}&=\sum_{n=1}^N\left[\frac{\partial F}{\partial\boldsymbol{\psi}_{n}}\cdot\delta\boldsymbol{\psi}_{n}+\delta\boldsymbol{\psi}_{n}^{\dagger}\cdot\frac{\partial F}{\partial\boldsymbol{\psi}_{n}^{\dagger}}\right]\nonumber\\
    &=2{\rm Re}\left[\sum_{n=1}^N\frac{\partial F}{\partial\boldsymbol{\psi}_{n}}\cdot\delta\boldsymbol{\psi}_{n}\right],\label{eq:deltaF_der}
\end{align}
where here we have applied the product rule to the general form of the objective function. The second line follows from the fact that the two added terms inside the square brackets on the right-hand-side of the first line are complex conjugates, and $a+a^*=2\text{Re}(a)$ for arbitrary complex number $a$. Substituting our expression for $\delta\boldsymbol{\psi}_{n}$ into Eqn.~\ref{eq:deltaF_der}, we can write the change in the objective function as
\begin{equation}
    \delta F_{\text{T}}=2{\rm Re}\left[\sum_{n=1}^N\left(\frac{\partial F}{\partial\boldsymbol{\psi}_{n}}\cdot\VecS_{>}\right)\cdot\delta\VecP_{n}\cdot\bigg(\VecS_{<}\cdot\boldsymbol{\chi}_{n}\bigg)\right]>0,\label{eq:Fdesign}
\end{equation}
which we demand to be positive -- i.e.\ the change improves the design of the $m^{\rm th}$ phase mask. Equation~\ref{eq:Fdesign} has a simple interpretation: To find the change in the phase mask to improve the transformation of input mode $n$ we must; (i) take the $n^{\rm th}$ input field $\boldsymbol{\chi}_{n}$ and propagate it up to (but not through) the boundary of the $m^{\rm th}$ phase mask using $\Vecv_{n,<}=\VecS_{<}\cdot\boldsymbol{\chi}_{n}$; and (ii) take the derivative of the objective function with respect to the $n^{\rm th}$ output field, and back--propagate that up to the other boundary of the $m^{\rm th}$ phase mask using $\Vecv_{n,>}=\partial F/\partial\boldsymbol{\psi}_{n}\cdot\VecS_{>}$. Note that, since the component matrices of the transfer matrix are unitary, we can write  $\partial F/\partial\boldsymbol{\psi}_{n}\cdot\VecS_{>}=(\VecS_{>}^{\dagger})^{\star}\cdot\partial F/\partial\boldsymbol{\psi}_{n}=(\VecS_{>}^{-1})^{\star}\cdot\partial F/\partial\boldsymbol{\psi}_{n}$.  The back propagation is thus the complex conjugate of the inverse of $\VecS_{>}$, which is equivalent to propagating through the $M-m$ planes in reverse order.  Given these two functions ($\Vecv_{n,<}$ and $\Vecv_{n,>}$) now defined on the two boundaries of the $m^{\rm th}$ phase mask, we can rewrite the change in the objective function (Eqn.~\ref{eq:Fdesign}) as
\begin{align}
    \delta F_{\text{T}}&=2\sum_{p=1}^P{\rm Re}\left[\left(\delta\VecP_{m}\right)_{p,p}\sum_{n=1}^N\left(\Vecv_{n,<}\right)_{p}\left(\Vecv_{n,>}\right)_{p}\right]\nonumber\\
    &=-2\sum_{p=1}^P\delta\theta_{m,p} {\rm Im}\left[{\rm e}^{{\rm i}\theta_{m,p}}\sum_{n=1}^N\left(\Vecv_{n,<}\right)_{p}\left(\Vecv_{n,>}\right)_{p}\right],\label{eq:deltaF_final}
\end{align}
where here we have now explicitly written out the matrix multiplications in the square brackets of Eqn.~\ref{eq:Fdesign} as a sum over all $P$ pixels, and $\left(\Vecv_{n,<}\right)_p$ labels element $p$ of $\Vecv_{n,<}$. The phase of pixel $p$ on plane $m$ is given by $\theta_{m,p}$, and the change in its phase value is given by $\delta\theta_{m,p}$. The second line of Eqn.~\ref{eq:deltaF_final} follows from the transfer matrix of the $m^{\rm th}$ phase mask being given by $\VecP_{m}={\rm diag}[{\rm e}^{i\theta_{m,p}}]$, and so
\begin{align}
\VecP_{m}+\delta\VecP_{m}&={\rm diag}[{\rm e}^{i(\theta_{m,p}+\delta\theta_{m,p})}]\nonumber\\
&\approx{\rm diag}[(1+i\delta\theta_{m,p}){\rm e}^{i(\theta_{m,p})}],
\end{align}
where here the approximation made is a truncated Taylor expansion of the first line. Therefore
\begin{align}
    \delta\VecP_{m} &= (\VecP_{m}+\delta\VecP_{m}) - \VecP_{m}\nonumber\\ &= {\rm diag}[i\delta\theta_{m,p}{\rm e}^{i\theta_{m,p}}],
\end{align}
and $(\delta\VecP_{m})_{p,p}$ labels the diagonal element of $\delta\VecP_{m}$ located at row and column $p$. We have also made use of the fact that for an arbitrary complex number $a$, ${\rm Re}(ia) = -{\rm Im}(a)$.

For the change in the objective function (\ref{eq:deltaF_final}) to be positive we therefore need to choose the phase change of each pixel $p$ such that
\begin{equation}
    {\rm sign}[\delta\theta_{m,p}]=-{\rm sign}\left\{{\rm Im}\left[{\rm e}^{{\rm i}\theta_{m,p}}\sum_{n=1}^N\left(\Vecv_{n,<}\right)_{p}\left(\Vecv_{n,>}\right)_{p}\right]\right\}.\label{eq:phase_design}
\end{equation}
The above prescription (Eqn.~\ref{eq:phase_design}) tells us how to independently change all $P$ pixels on a phase plane in one computation -- the phase value of each pixel $p$ is either increased or decreased by a small fixed amount (the step size). This parallel gradient computation is common to all such adjoint optimization approaches where we work in terms of both a forwards ($\Vecv_{n,<}$) and backwards ($\Vecv_{n,>}$) formulation of the problem. The optimisation is then iteratively repeated until convergence -- i.e. arriving at a solution equivalent to a local maximum.\\

\noindent{\bf \S3: Gradient ascent objective functions}\\\label{SI_obj}
\noindent{\it Mode Sorting}: The objective function used to design the mode sorters in this work is given by $F_T = \sum_{n=1}^NF_n$, where
\begin{equation}\label{Eqn:objectiveSI}
    F_n = \underbrace{\alpha \left|\boldsymbol{\psi}^{\dagger}_n\cdot\boldsymbol{\phi}_n\right|^2}_\text{Fidelity - phase free} -\underbrace{\beta\;{\rm Re}\left[\boldsymbol{\psi}^{\dagger}_n\cdot\boldsymbol{\psi}^{\text{cr}}_n\right]}_\text{Cross-talk} +\underbrace{\gamma\;{\rm Re}\left[\boldsymbol{\psi}^{\dagger}_n\cdot\boldsymbol{\psi}^{\text{bk}}_n\right]}_\text{Efficiency}.
\end{equation}
In this case the gradient of $F_n$ with respect to $\boldsymbol{\psi}_{n}$ is given by 
\begin{equation}
    \frac{\partial F_{n}}{\partial\boldsymbol{\psi}_{n}}=\alpha\left(\boldsymbol{\psi}^{\dagger}_{n}\cdot\boldsymbol{\phi}_{n}\right)\boldsymbol{\phi}_{n}^{\dagger}-\tfrac{1}{2}\beta\boldsymbol{\psi}_{n}^{{\rm cr}\;\dagger}+\tfrac{1}{2}\gamma\boldsymbol{\psi}_{n}^{{\rm bk}\;\dagger}.\label{Eqn:objectiveGradSI}
\end{equation}
To differentiate the fidelity term of Eqn.~\ref{Eqn:objectiveSI}, we have used the fact that for arbitrary complex column vectors $\boldsymbol{a}$ and $\boldsymbol{b}$, of the same length, $\left|\boldsymbol{a}^{\dagger}\cdot\boldsymbol{b}\right|^2 = \left(\boldsymbol{a}^{\dagger}\cdot\boldsymbol{b}\right)\left(\boldsymbol{b}^{\dagger}\cdot\boldsymbol{a}\right)$. For the other terms we can make use of the identity ${\rm Re}\left(\boldsymbol{a}^{\dagger}\cdot\boldsymbol{b}\right) = \tfrac{1}{2}\left[\left(\boldsymbol{a}^{\dagger}\cdot\boldsymbol{b}\right)+\left(\boldsymbol{b}^{\dagger}\cdot\boldsymbol{a}\right)\right]$ to differentiate them. The vectors $\boldsymbol{\psi}_{n}$ and $\boldsymbol{\psi}_{n}^\dagger$ are treated as independent variables, which accounts for their complex nature when deriving the differentials.\\

\noindent{\it Optical circuits}: The objective function used to design the optical circuits in our work is given by $F_T = \sum_{n=1}^NF_n$, where
\begin{equation}\label{Eqn:objective2SI}
    F_n = \underbrace{\alpha {\rm Re}\left[\boldsymbol{\psi}^{\dagger}_n\cdot\boldsymbol{\phi}_n)\right]}_\text{Fidelity - phase fixed}+\underbrace{\gamma\;{\rm Re}\left[\boldsymbol{\psi}^{\dagger}_n\cdot\boldsymbol{\psi}^{\text{bk}}_n\right]}_\text{Efficiency}.
\end{equation}
In this case the gradient of $F_n$ with respect to $\boldsymbol{\psi}_{n}$ is given by 
\begin{equation}
    \frac{\partial F_{n}}{\partial\boldsymbol{\psi}_{n}}=\tfrac{1}{2}\left(\alpha\boldsymbol{\phi}_{n}^{\dagger}+\gamma\boldsymbol{\psi}_{n}^{{\rm bk}\;\dagger}\right).\label{Eqn:objectiveGradSI}
\end{equation}
\\

\noindent{\bf \S4 Relationship between gradient ascent and the wavefront matching method}\\
The wavefront matching method is an adjoint inverse design algorithm that can be used to design MPLCs. The WMM optimises the overlap between the target output modes (labelled by index $n$) $\boldsymbol{\phi}_n$, and the actual output spatial modes $\boldsymbol{\psi}_n$. The objective function to optimise is given by
\begin{equation}\label{WMMobjectivephasefixed}
    F_T = \left|\sum_{n=1}^N\big(\boldsymbol{\phi}^{\dagger}_n\cdot\boldsymbol{\psi}_n\big)\right|^2,
\end{equation}
which enforces the relative output phase between modes. In this special case, the optimum change to the phase of all pixels $p$ on phase plane $m$ can be directly calculated by expanding $F_T$ as follows
\begin{align}\label{Eq:WMMSI}
    F_T &= \left|\sum_{n=1}^N\big(\boldsymbol{\phi}^{\dagger}_n\cdot\boldsymbol{\psi}_n\big)\right|^2\nonumber\\
    &= \left|\sum_{n=1}^N\big(\boldsymbol{\phi}^{\dagger}_n\cdot\VecS\cdot\boldsymbol{\chi}_n\big)\right|^2\nonumber\\
    &= \left|\sum_{n=1}^N\big(\Vecv_{n,>}\cdot\VecP_m\cdot\Vecv_{n,<}\big)\right|^2\nonumber\\
    &= \left|\sum_{n=1}^N\sum_{p=1}^P\big(\left(\Vecv_{n,>}\right)_p{\rm e}^{i\theta_{m,p}}\left(\Vecv_{n,<}\right)_p\big)\right|^2\nonumber\\
    &= \left|\sum_{p=1}^P{\rm e}^{i\theta_{m,p}}\sum_{n=1}^N\big(\left(\Vecv_{n,>}\right)_p\left(\Vecv_{n,<}\right)_p\big)\right|^2.
\end{align}
The second line of Eqn.~\ref{Eq:WMMSI} is obtained by writing the output field $\boldsymbol{\psi}_n$ in terms of the input mode $\boldsymbol{\chi}_n$ and the transfer matrix of the MPLC $\VecS$, i.e.\ $\boldsymbol{\psi}_n = \VecS\cdot\boldsymbol{\chi}_n$. Next we factorise $\VecS$ using main text Eqn.~\ref{Eqn:MPLCreal} to single out the action of the $m^{\text{th}}$ phase plane ($\VecP_m$), yielding line three. Here $\Vecv_{n,<} = \VecS_{<}\cdot\boldsymbol{\chi}_{n}$ (using the same nomenclature as introduced in \S2.), which represents the propagation of input mode $\boldsymbol{\chi}_{n}$ up to the front face of the $m^{\text{th}}$ phase plane. $\Vecv_{n,>} =\boldsymbol{\phi}_{n}^\dagger\cdot\VecS_{>}$ represents the backward-propagation of the target output mode $\boldsymbol{\psi}_{n}$ up to the back face of the $m^{\text{th}}$ phase plane. In line four we have rewritten the matrix multiplication as a sum over all pixels $p$, and then rearranged this in line five -- where ${\rm e}^{i\theta_{m,p}}$ can be taken outside the sum over $n$ as it is independent of $n$. Finally, we can see that $F_n$ takes a maximal value when the sum over $p$ is real and positive, i.e.\ when the phase of pixel $p$ of plane $m$, $\theta_{m,p}$, is chosen to equal the difference between the phase of elements $\left(\Vecv_{n,>}\right)_p$ and $\left(\Vecv_{n,<}\right)_p$, averaged over all $N$ modes, i.e.\
\begin{equation}
    \theta_{m,p} = -\arg\left[\sum_{n=1}^N\big(\left(\Vecv_{n,>}\right)_p\left(\Vecv_{n,<}\right)_p\big)\right].
\end{equation}
The optimal new phase values of all $P$ pixels on plane $m$ can be calculated simultaneously in this manner, and the WMM then proceeds by iterating repeatedly over all $M$ planes until convergence.

We can see the WMM involves similar forward and backward field propagation steps to our gradient ascent optimiser, as it is another form of adjoint optimisation. However, rather than relying on a gradient calculation $\delta F_T$ followed by incremental modifications to each phase plane, WMM instead enables direct calculation of larger spatially varying changes to the phase profile of each plane. Therefore the WMM tends to converge more quickly than our gradient ascent approach. We highlight that this direct calculation of the optimum change in phase of each pixel is only possible for objective functions based on optimizing the overlap integral of a set of output and target spatial modes.
Therefore, using WMM to optimise the objective function given in Eqn.~\ref{WMMobjectivephasefixed} is equivalent to using our gradient ascent approach with the objective function given in Eqn.\ref{Eqn:objective2SI}, with $\alpha=1$; $\gamma = 0$.

Finally, we note that modal weighting and a smoothing constraint may also be included in the WMM, which modifies the choice of optimal phase to
\begin{equation}
    \theta_{m,p} = -\arg\left[\sum_{n=1}^N\big(a_n\left(\Vecv_{n,>}\right)_p\left(\Vecv_{n,<}\right)_p+b\big)\right].
\end{equation}
Here $a_n$ is a mode dependent weight that can be used to prioritise the importance of sorting some modes compared to others. It is useful to even out the mode sorting efficiency if necessary.
Scalar $b$ is a real-valued adjustable parameter first suggested in~\cite{fontaine2019laguerre}, that allows one to suppress the onset of noise in the regions of the mask with low incident light intensity and consequently poorly defined phase values in functions $\VecV_{n,>}$ and $\VecV_{n,<}$. It is also possible to increase $b$ to encourage the optimised phase masks to be more smoothly-varying, at a cost of reducing the overall transformation performance.\\

\noindent{\bf \S5: Quantification of mode sorter performance}\\ \label{SI_perf_quant}
We use three parameters to characterize the performance of each mode sorter or optical circuit shown in our work: {\it average cross-talk}, {\it average efficiency}, and {\it average localized fidelity}, which we define below.

For mode sorters, we define output channels (shown in Fig.~\ref{Fig:concept}(b)) as circular areas around every Gaussian spot in the output plane. For every input mode we then calculate a fraction of the total power in every output channel and arrange the results into a column vector. Having $N$ input-output pairs of modes would result in constructing a $N$$\cross$$N$ cross-talk matrix, like shown in Fig. \ref{Fig:setup_speckle} (c, d). The off-diagonal elements of this matrix show the amount of light entering the wrong output channels in the output plane. So in an ideal case with no cross-talk in the system, the cross-talk matrix would be equal to the identity matrix. The cross-talk $c_n$ of every mode is given by the ratio of the sum of the power in all the wrong output channels to the sum of the power in all the output channels:
\begin{equation} \label{Eqn:crs}
    c_n = \frac{\displaystyle\sum_{i=1}^{N}\left(\displaystyle\iint \left|\psi_n(\vec{r}) \right|^2 {\rm d}s_i(\vec{r})\right) - \iint \left|\psi_n(\vec{r}) \right|^2 {\rm d}s_n(\vec{r}) }{\displaystyle\sum_{i=1}^{N}\displaystyle\iint \left|\psi_n(\vec{r}) \right|^2 {\rm d}s_i(\vec{r})} = 1 - \frac{\displaystyle\iint \left|\psi_n(\vec{r}) \right|^2 {\rm d}s_n(\vec{r}) }{\displaystyle\sum_{i=1}^{N}\displaystyle\iint \left|\psi_n(\vec{r}) \right|^2 {\rm d}s_i(\vec{r})},
\end{equation}
where $\psi_n$ is the field of the $n^{\text{th}}$ output mode, ${\rm d}s_n$ is an area element drown from just the $n^{\text{th}}$ target output channel, ${\rm d}s$ is an area element drawn from the entire output plane, and $\vec{r}$ is the coordinate vector. $c_n$ is then averaged over all modes to calculate the average cross-talk $C_{\text{r}}$.

\begin{figure}[t]
   \includegraphics[width=1.0\columnwidth]{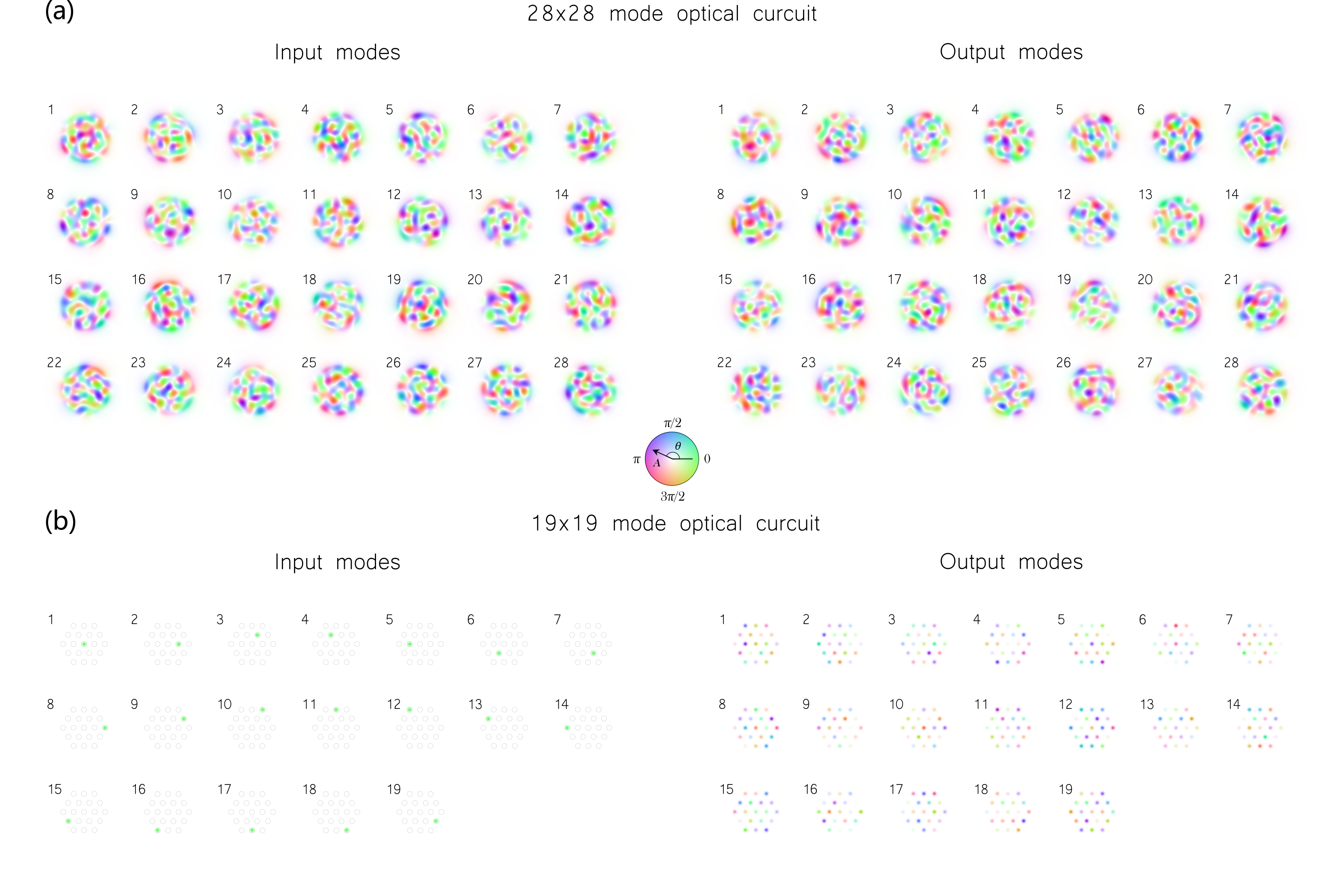}
   \caption{(a) Inputs and outputs for the 5-plane 28-mode arbitrary speckle transformer. (b) Inputs and outputs for the 5-plane 19-mode beam combiner.}
   \label{Fig:suppl_opt_c}
\end{figure}

The efficiency of every mode, $e_n$ is calculated by dividing the power in the target output channel by the total power in the input mode: 
\begin{equation} \label{Eqn:eff}
    e_n = \frac{\displaystyle\iint \left|\psi_n(\vec{r}) \right|^2 {\rm d}s_n(\vec{r})}{\displaystyle\iint \left|\chi_n(\vec{r}) \right|^2 {\rm d}s(\vec{r})},
\end{equation}
where $\chi_n$ is the complex field of the $n^{\text{th}}$ input mode. In our experiments we found the efficiency to be significantly lower than it is in our simulations due to the losses introduced by a limited reflectivity of our LCSLM screen -- which was a model not equipped with a dielectric back-plane which can be used to promote high levels of reflection efficiency. Efficiency can be improved in the future by using LCSLMs with such a dielectric back-plane, or by using a lithographically etched set of phase masks instead of an SLM as part of the MPLC system.

We also define a parameter to quantify the fidelity of the output modes -- which we term the localized fidelity. The localized fidelity of the $n^{\text{th}}$ output mode is given by
 \begin{equation} \label{Eqn:loc_fid}
    f_n = \displaystyle\iint \left|\psi_n(\vec{r}) \cdot \phi^*_n(\vec{r})\right|^2 {\rm d}s_n(\vec{r})
\end{equation}
Here the term {\it localised} refers to the fact that the normalisation of $\psi_n$ is only carried out over the extent of the output channel of interest (i.e.\ integrating only over area elements ${\rm d}s_n(\vec{r})$) -- thus removing the contribution of any power scattered outside the output channel.
 
For optical circuits, we define the target output channel(s) $s_n$ as shown in Fig.~\ref{Fig:OptCircuit}(d,i), and use the above-mentioned average efficiency and average localized fidelity to quantify the performance. The quality of an optical circuit can also be quantified by calculating the fidelity of the engineered matrix transform itself. This is given by \begin{equation}
    f_{\text{M}} = \left|\sum_{i=1}^N\sum_{j=1}^N\VecM^*_{ij}\VecM'_{ij}\right|^2,
\end{equation}
where $\VecM$ is the normalised target matrix transform (shown in the insets of Figs.~\ref{Fig:OptCircuit}(e,j)), $\VecM'$ is the normalised actual transform accomplished by the MPLC. $f_{\text{M}}$ differs from the average output fidelity $f$, as $f_{\text{M}}$ ignores the presence of any transmitted background speckle in spatial modes outside of the output basis.\\
 
 \noindent{\bf \S6: Input and output modes for optical circuit designs}\\
Figure~\ref{Fig:suppl_opt_c} show the input and output modes for the two optical circuit designs shown in main text Fig.~3.\\
 
\noindent{\bf \S7: Experimental optimisation of misalignements when implementing an MPLC}\\ \label{SI_exp_optim}
As discussed in the main text, experimental realisation of high-dimensional mode sorters, which in general require the use of complicated non-smoothly varying phase masks, involves dealing with $\sim 2M+4$ degrees of freedom (DoF) of the physical system. These degrees of freedom are: degree of defocus (1 DoF) and size of the beam waist (1 DoF) of the incoming beams, the distance (1 DoF) between the SLM and the adjacent mirror, the lateral location of each phase mask ($2\times M$ DoF, where $M$ is the number of phase masks used), and the distance of the output plane (1 DoF). 
 
However, the number of DoFs that need to be taken into consideration in practice may be even larger. For example, if the mirror facing the LCSLM is placed not perfectly parallel to the LCSLM, the distance light travels between the each pair of phase masks may vary. Flatness imperfections of the SLM screen would also contribute to these distance differences. Thus, one may choose to optimize not only a single distance between all pairs of phase masks, but $(M - 1)$ distances -- different for every pair of masks. One of the other potential degrees of freedom to optimise is the angle of incidence of the incoming beam. When input modes arrive to the first MPLC plane at a high ($ >\ang{5}$) angle, the transverse cross-section approximation of the modes used in the phase mask design algorithm may no longer be accurate enough, so one could consider generating input modes for the design algorithm in an inclined plane that reflects what happens in experiment, and, potentially, optimise this parameter.
 
Although optimising all of the degrees of freedom is essential to reach the simulated performance in experiment, we find that it is possible to drastically minimize the measured cross-talk by performing a pixel-perfect alignment of the lateral positions of phase masks on the SLM screen. While it is possible to precisely measure the angle of incidence of the incoming into the MPLC beam, and it is possible to put a mirror precisely in parallel to the SLM screen using custom-made mechanical elements, it is not possible to have control over the non-flat nature of the SLM screen. This means that (to the first order of aberration) the beam reflected from each phase mask may obtain a small angular deviation. In such a way flatness imperfections of the SLM screen can contribute to the misalignment of masks' lateral positions on a scale of tens of microns. Hence, individually tuning the lateral location of each mask becomes necessary. 
 
As phase masks in experiment can only be moved an integer number of pixels aside on the SLM screen, we develop an optimisation procedure that involves using an integer genetic algorithm to mitigate the misalignments by tweaking the lateral position of each mask by a few pixels. We use the genetic algorithm optimisation in a feedback loop to find the best-performing positions. First, we display phase masks on the SLM screen in experiment using initial lateral positions approximated during the manual alignment. After this, a set of $N$ input modes are projected into the MPLC and a camera captures a set of $N$ intensity profiles of the outputs. Then the average cross-talk value between the output channels is calculated and sent as the feedback to the genetic algorithm. After receiving the feedback, genetic algorithm generates a new guess (or a generation of guesses) to try using it as new lateral coordinates (in pixels), after which the process repeats.
 
Using the described technique, it is possible to optimise not only lateral positions, but also the distance between phase masks (axial positions), and degrees of defocus on the input and on the output of the MPLC. This can be done by generating sets of phase mask designs beforehand, each one for a slightly different spacing between planes, and for different distances to the input and output planes to simulate defocus. Then, after specifying the search limits for each parameter, the genetic algorithm cycles through the axial and lateral positions of the masks in experiment -- by updating sets of masks on the SLM screen and moving them a few pixels aside.

Running this optimisation for a 5-plane setup (so $5\times2 + 3 = 13$ degrees of freedom, including optimisation of the spacing between planes, and degrees of defocus on the input and output of the MPLC) for $10$ generations in experiment, which is usually more than enough to find a local minimum, takes $\sim30$ minutes in our implementation (Matlab and LabView). Also, in order to optimize axial positions and defocus, a few hundreds of different sets of masks have to be designed in advance, which may usually take a few hours depending on a number of sets to design and can be done overnight prior to the experimental optimisation.
 
Searching for the optimal beam waist of the incoming into the MPLC beam can also be incorporated into the alignment algorithm by designing extra sets of masks beforehand, each for slightly different input beam waist. Then, keeping the beam waist of the projected into the MPLC beam constant in experiment, the genetic algorithm would run through the different sets of masks designed for different beam waists to locate the optimal value out of the set.\\
 
\noindent{\bf \S8: Detailed experimental mode sorting results}\\ \label{SI_exp_results}
In this supplementary section we show additional detail of the obtained experimental results, including mode sorting using both 3-plane and 5-plane MPLCs. Each experiment here was conducted using the Hamamatsu X13138-01 SLM (without a high reflectivity dielectric back-plane) with a spatial resolution of $1280\times1024$ pixels. The width of the SLM screen in pixels constrains spatial resolution of each phase mask for a fixed number of masks used. Thus, in experiments shown below, in 3-plane MPLC systems we use $400\times400$ pix phase masks, and in 5-plane MPLCs we use $400\times200$ pix masks.
Based on our understanding of the symmetry of presented below modal sets and how difficult it is to sort certain sets, sub-sections here show mode sorters of increasing complexity.

We use average cross-talk to quantify the performance of each mode sorter in experiment, chosen as it can be calculated directly from camera images of the intensity of the outputs, without the need for interferometric full-field measurements. see SI \S5 for the definition of the average cross-talk. We note that the obtained values of cross-talk depend weakly on the radius of the target output channel we define in an output plane of the MPLC.\\

\begin{figure}
  \includegraphics[width=0.7\columnwidth]{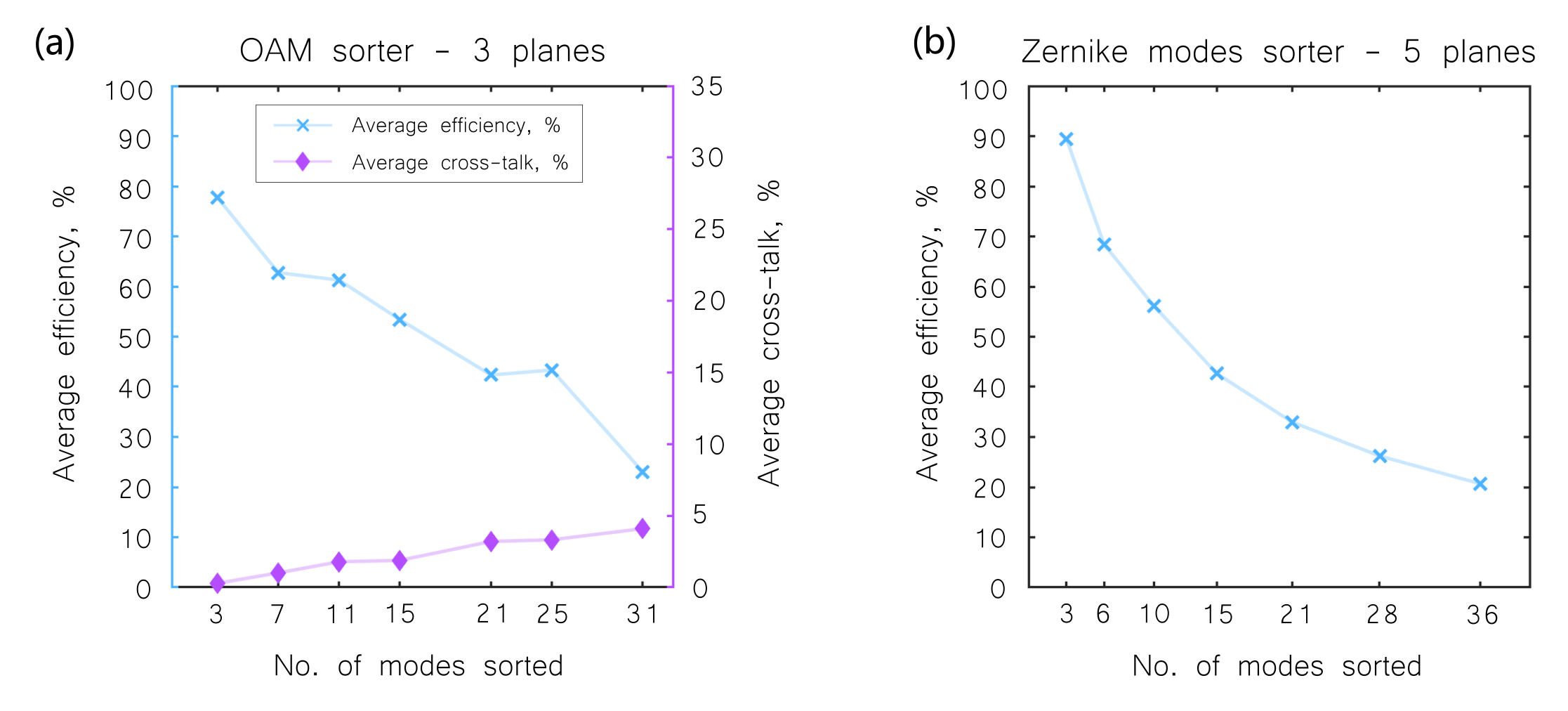}
  \caption{Mode sorter performance simulated using phase masks designed with the gradient ascent algorithm, resolution of each phase mask is $700\cross700$ pix. (a) Average efficiency and average cross-talk of a 3-plane OAM sorter; (b) Average efficiency of a 5-plane Zernike mode sorter as a function of number of modes sorted.}
  \label{Fig:suppl_eff}
\end{figure}

\noindent{\it Sorting of orbital angular momentum states}\\
In addition to the experiment results of OAM sorting shown in main text Fig.~\ref{Fig:oam}, we also include supplementary Fig.~\ref{Fig:suppl_eff}(a), which demonstrates the simulated performance of a 3-plane OAM sorter designed using gradient ascent optimisation.\\

\begin{figure}
   \includegraphics[width=1.0\columnwidth]{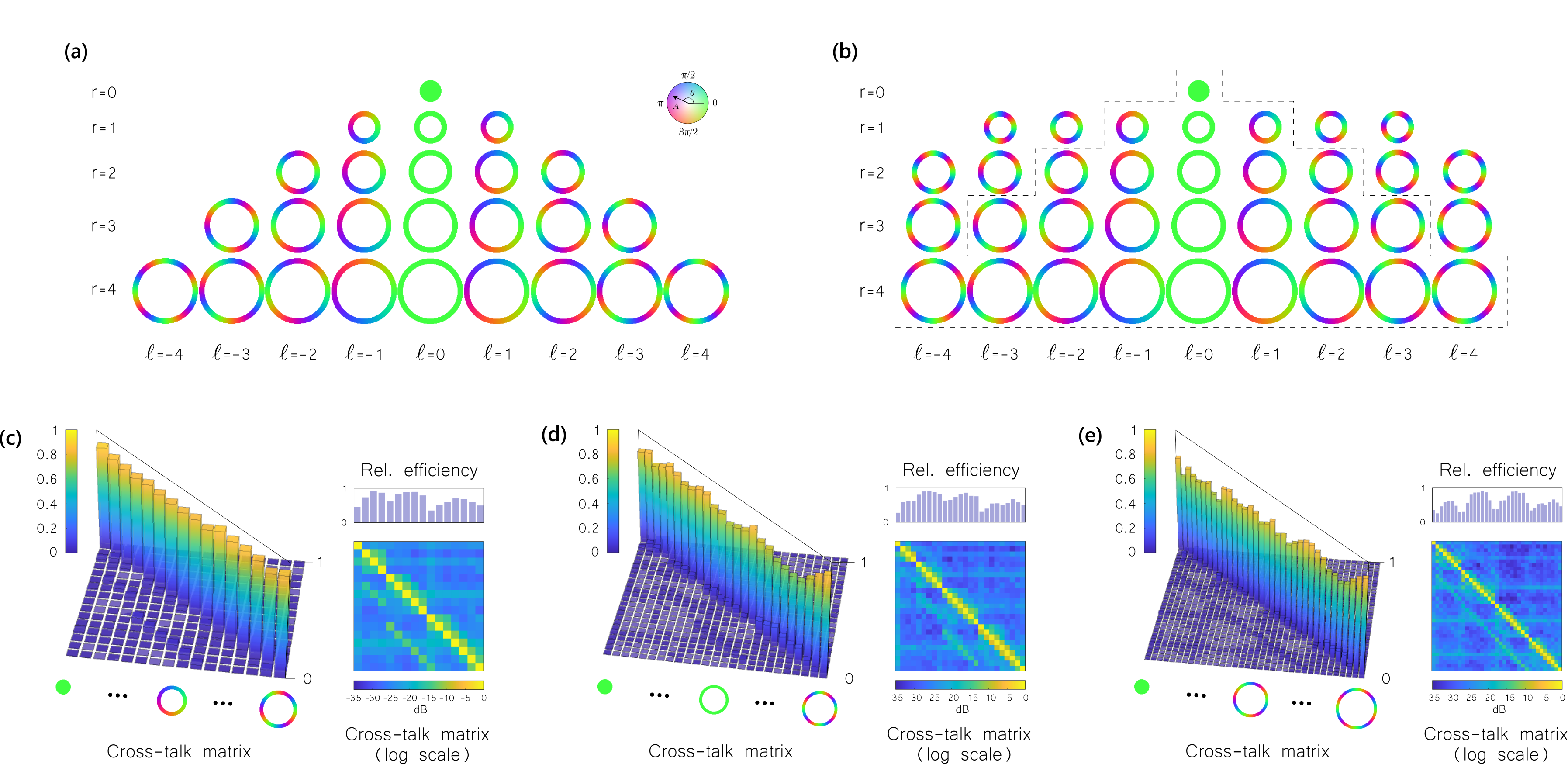}
   \caption{(a) An example of the basis containing 25 annuli that carry OAM. Modes are divided into groups based on their radial index $r$, with a certain integer number of groups being sorted in experiment. (b) An example of the extended basis containing 35 annuli, sorting of which was demonstrated in experiment using phase masks designed to sort only 25 annuli. (c-e) Experimental results for the 3-plane sorter of annuli carrying OAM. Average cross-talk $C_{\text{r}}$ for 16, 25 and 35 modes sorted respectively: $\sim$ 15.6\%; 21.7\%; 21.6\%.}
   \label{Fig:suppl_bess_an}
\end{figure}

\noindent{\it Sorting of annuli carrying OAM}\\
Conventional OAM sorters are usually designed to sort only azimuthal component of an incoming beam. Here we consider a task of sorting radial and azimuthal components at the same time - which is more challenging. We start with what we consider to be the simplest way to define radial spatial modes, and generate annuli divided into different groups by their radial component index $r$, and apply helical phase to them in such a way that every next radial group carries two additional OAM states. The Fourier Transform of these functions approximate the basis of Bessel functions -- with their radial extent modulated by a {\it sinc} function. However, when represented as annuli, it is important to note that intensity profiles of any two modes from different radial groups do not overlap at all, while intensity profiles of any two modes from the same radial group have exactly the same shape. Thus the radial modes are already spatially separated, and the task of the sorter is to reshape these modes into Gaussian spots, while simultaneously spatially separating the spatially overlapping azimuthal modes. Figure \ref{Fig:suppl_bess_an}(a) demonstrates an example of the basis containing 25 annuli that carry OAM. Figure \ref{Fig:suppl_bess_an}(c-e) shows the experimental results for the 3-plane sorter of annuli carrying OAM.

Figure \ref{Fig:suppl_bess_an}(e) shows the experimental result of sorting up to 35 annuli carrying OAM. This particular result is based on an interesting phenomenon of phase masks inheriting properties of a set of modes $N$ and being able to sort even a larger set $>N$ of modes. Figure \ref{Fig:suppl_bess_an}(b) shows the original set of 25 modes (inside the dashed line) phase masks were designed for, and also the additional 10 modes (outside the dashed line) this set of masks was able to sort in experiment.

It is also useful to compare the experimental results of sorting annuli carrying OAM to the performance simulated using different design algorithms and different resolution of the field-of-view in pixels. As can be seen in Fig.\ref{Fig:suppl_bess_an}(d), average cross-talk $C_{\text{r}}$ of the experimental 25-mode sorter using 3 phase masks ($400\times400$ pix. resolution each) is $\sim 21.7\%$. It is larger than the simulated performance of the system with the same number of phase masks ($400\times400$ pix. each) designed by the wavefront matching method ($\sim 4.5\%$) due to the experimental misalignments and inter-pixel cross-talk of the SLM itself \cite{moser2019model, pushkina2020comprehensive}. While the lowest average cross-talk possible ($\sim 3.5\%$) is achievable in simulation using our gradient ascent algorithm with a $700\times700$ pix. resolution of each phase mask.\\

\noindent{\it Bessel beam sorting}\\
In the previous sub-section we demonstrated sorting of the modes with radial and azimuthal components, however intensity profiles of any two spatial modes from different radial groups in those sets were not overlapping. Now we demonstrate what we consider to be a more challenging task -- Bessel beam sorting. Bessel beams also carry radial and azimuthal components, but any two modes from two different radial groups overlap quite a lot. In other words, we consider the task of sorting a certain modal set to be generally more challenging for sets of beams that are more spatially overlapped at the input.

\begin{figure}
   \includegraphics[width=0.8\columnwidth]{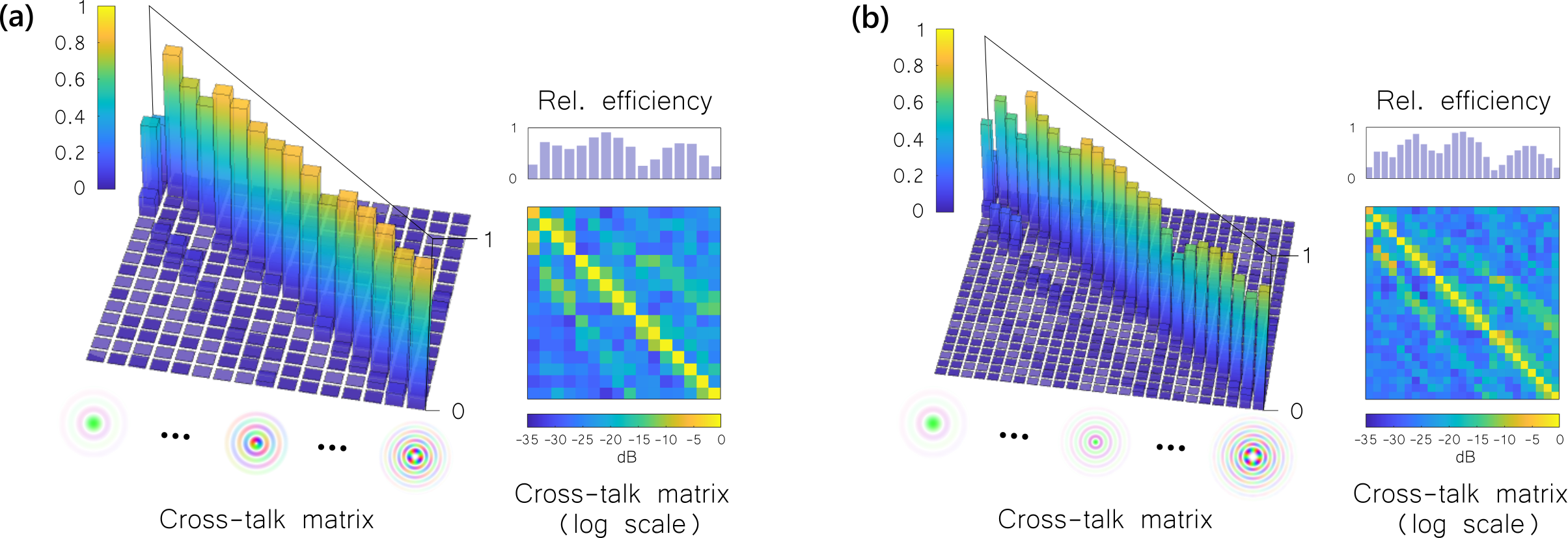}
   \caption{Experimental results for the 3-plane Bessel beam sorter. Bessel beams are divided into groups based on their radial index $r$, with a certain integer number of groups being sorted in experiment. Average cross-talk $C_{\text{r}}$ for 16 and 25 modes sorted (a, b): $\sim$ 20.7\%; 30.5\%.}
   \label{Fig:suppl_bess_b}
\end{figure}

Figure \ref{Fig:suppl_bess_b} shows the experimental results for the 3-plane Bessel beam sorter. We also compare the experimental results of sorting Bessel beams to the performance simulated using different design algorithms and different resolution of the field of view in pixels. As shown in Fig.~\ref{Fig:suppl_bess_b}(b), average cross-talk $C_{\text{r}}$ of the experimental 25-mode sorter using 3 phase masks ($400\times400$ pix. resolution each) is $\sim 30.5\%$. It is larger than the simulated performance of the same system with the phase masks ($400\times400$ pix. each) designed by the wavefront matching method ($\sim 12.5\%$) due to the same reasons described in a previous section. It is possible to achieve much lower levels of average cross-talk ($\sim 3.0\%$) in simulation when sorting 25 Bessel beams with 3 phase planes using our gradient ascent algorithm with a $700\times700$ pix. resolution of each phase mask.\\

\noindent{\it Zernike mode sorting}\\
The Zernike modes $Z$ are a set of orthogonal functions which are typically associated with low order aberrations in optical systems:
\begin{equation}
    Z^{m}_n(\rho,\phi) = \sum_{k=0}^{\frac{n-m}{2}}\frac{(-1)^{k}(n-k)!}{k!(\frac{n+m}{2} - k)! (\frac{n-m}{2} - k)!} \rho^{n-2k} \cross
    \begin{cases}
      \cos(m\phi) & \text{if $m \geq 0$}\\
      \sin(m\phi) & \text{if $m < 0$},
    \end{cases}
    \label{eq:zern}
\end{equation}
where $\rho$ and $\phi$ are the radial and azimuthal coordinates respectively, and $m$ and $n$ specify the mode index.

\begin{figure}
   \includegraphics[width=1.0\columnwidth]{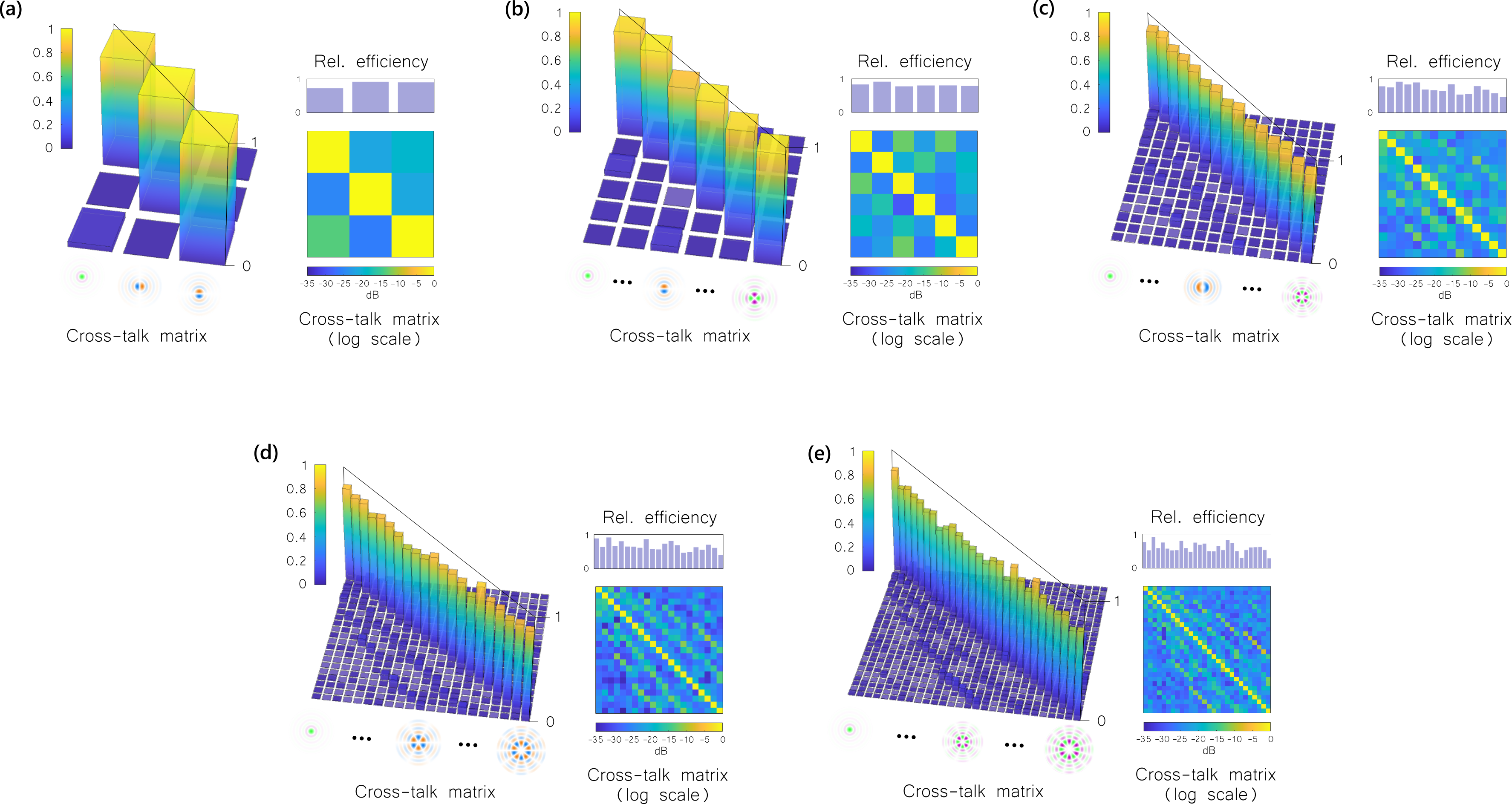}
   \caption{Experimental results for the 5-plane Zernike mode sorter, in addition to the ones demonstrated in Fig.~\ref{Fig:Zern}. Average cross-talk $C_{\text{r}}$ for 3, 6, 15, 21 and 28 modes sorted respectively (a-e): $\sim$ 2.5\%; 6.9\%; 16.7\%; 19.1\%; 26.6\%.}
   \label{Fig:suppl_zern}
\end{figure}

Next, we show the detailed experimental results for the each data point of the graph demonstrated in Fig.~\ref{Fig:Zern}(b). Figure \ref{Fig:suppl_zern} shows the experimental results for a series of 5-plane Zernike sorters, additional to the ones already demonstrated in Fig.~\ref{Fig:Zern}. We define Zernike modes incident on the first plane of the mode sorter as the Fourier transform of the function shown in Eqn.~\ref{eq:zern}. We then divide them into groups based on their index $n$ and sort a certain integer number of groups in experiment, with each row of Gaussian spots on the output corresponding to each group. In other words, for each $N$ mode sorter shown here, we sort first $N$ Zernikes based on their index $j$ (OSA/ANSI notation), where $j = 0, ..., N-1$.

Figure \ref{Fig:suppl_eff}(b) also shows simulated average efficiency of such 5-plane Zernike sorter designed using gradient ascent optimisation. While average efficiency in experiment is heavily impacted by losses introduced at every plane of the MPLC, level of which depends on a particular SLM model.\\

\begin{figure}[h!]
   \includegraphics[width=1.0\columnwidth]{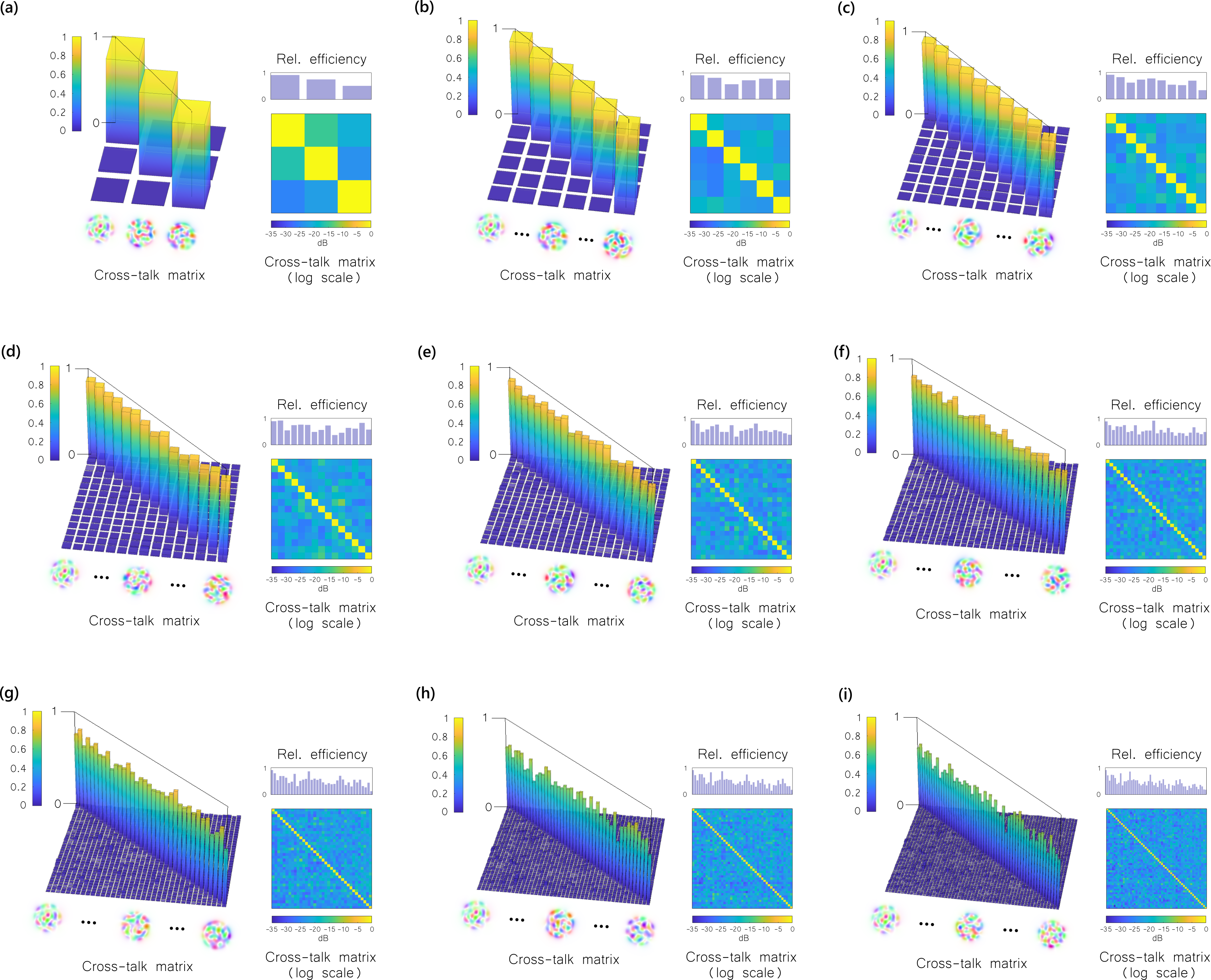}
   \caption{Experimental results for the 5-plane speckle 'mode' sorter. Average cross-talk $C_{\text{r}}$ for 3, 6, 10, 15, 21, 28, 36, 45 and 55 modes sorted respectively (a-i): $\sim$ 2.6\%; 4.1\%; 8.1\%; 12.9\%; 16.5\%; 20.9\%; 26.2\%; 34.8\%; 37.9\%.}
   \label{Fig:suppl_speckle}
\end{figure}

\noindent{\it Sorting of orthogonal speckle patterns}\\
We consider a task of sorting a set of randomly generated orthogonal speckles to be the most general case. Compared to all the other modal sets demonstrated in this section, a set of random speckles has no underlying symmetries with its amplitude and phase distribution. As mentioned earlier, in order to spatially localize random speckles, we generate them as the complex weighted sum of a set of orthogonal step-index multimode fibre modes. We simulate a multimode fibre with $NA = 0.01$ and a core radius $r = 160$\,$\mu$m that supports up to 65 eigenmodes at $\lambda = 633$\,nm. In this section, we demonstrate the detailed experimental results for the each data point of the graph shown in Fig.~\ref{Fig:setup_speckle}(b). Figure~\ref{Fig:suppl_speckle} shows the experimental results for a series of 5-plane speckle sorters.\\

\noindent{\bf \S9: Description of supplementary movies}\\
\noindent {\it Supplementary movie 1} shows the recorded outputs of a 5-plane, 55 speckle-mode sorter as it is sequentially illuminated with each input speckle pattern in turn. Simulated outputs are shown alongside with the recorded experimental outputs.\\
\noindent {\it Supplementary movie 2} shows the experimentally measured outputs of a 3-plane OAM sorter, that has been designed to sort 21 orbital angular momentum states of light. Laguerre-Gaussian modes with the radial index $p = 0$ are used as input modes to the system.\\
\noindent {\it Supplementary movie 3} shows the experimentally measured outputs of a 5-plane, 36 Zernike-mode sorter.

\end{document}